%Paper: hep-ph/9304207
%From: glin@phys.sinica.edu.tw (LIN guey-lin)
%Date: Sat, 3 Apr 93 18:00:53 CST
%Date (revised): Thu, 14 Oct 93 16:52:54 CST

\input phyzzx

%\magnification=1200
\baselineskip=20pt
%\theory
\nopubblock
\rightline{UM--TH--93--05}
\rightline{IP--ASTP--09--93}
\rightline{Freiburg THEP--93/02}
\rightline{hep-ph/9304207}
\rightline{16 September 1993}
\smallskip
\pubtype{}
\titlepage
\title{Heavy Top Quark Effects in Vertices with Light External
Fermions}
\author{Guey-Lin Lin}
\address{Institute of Physics, Academia Sinica, Taipei, Taiwan 11529,
Republic of China}
\andauthor{Herbert Steger}
\address{Fakult\"at f\"ur Physik,
Albert-Ludwigs-Universit\"at\nextline
D-7800 Freiburg im Breisgau, Germany}
\andauthor{York-Peng Yao}
\address{Department of Physics,
Randall Laboratory\nextline
University of Michigan\nextline
Ann Arbor, Michigan 48109-1120, USA}
\endpage
\abstract
\smallskip
We extend our previous work on non-linear realization of the top
quark field, which is a consequence of its being much heavier than
any other scales.  One loop effective Lagrangian to account for
virtual top effects is constructed in this article, for
processes where there are two external fermions, as well as any number of
bosons.
Particularly, we focus on those terms which have power dependence on
the top mass $m_t$.
\endpage
\noindent{\bf 1. INTRODUCTION}
\bigskip
There is heightened interest in top quark physics.  From various
theoretical and experimental input, it is expected that within the
standard model $m_t \approx 145 \ {\rm GeV}$ [1]. It may prove to be the most
massive quark and potentially can offer us some hint to the mysterious
mass generating mechanism of spontaneous symmetry breaking.
\smallskip
Of course, the most conclusive evidence to establish the existence
would be to produce naked tops.  This is eagerly anticipated to
happen in the Tevatron in the near future.  Even so, it seems that
just from statistics considerations, one still has to depend very
heavily on low energy physics to gain understanding of the parameters,
such as K-M matrix elements $V_{tj} \ (j=b,s,d)$, and the dynamics of this
new member of the third family.
\smallskip
We have been investigating physics related to the top, particularly in
the method of effective Lagrangians.
One emphasis of the present article is to show how one can ${\it read \ off}$
results from this paper and our previous works and apply them
to some interesting processes.  We want to convey to our readers the
simplicity and universality of our treatment.

For low energy processes, $p_{ext},m, m_{W,Z} << m_t$, the top field
is expressed in terms of the Higgs fields and the b field [2]
$$t_L=-{{\phi^+ b_L}\over {\phi^0}}, \ \ \ t_R=0. \eqno (1.1)$$
The symmetry $SU(2)_L\otimes U(1)_Y $ is realized non-linearly.  This
seems to be a general feature of a strong coupling theory; a heavy
Higgs is another example [3]. The aim is to condense and summarize the top
virtual effects in an effective
Lagrangian

$$L_{eff}=\sum_i C_iO_i, \eqno (1.2)$$
where $O_i$ is a set of local operators made of light fields and derivatives
and $C_i$ is the corresponding set of effective coefficients, which are
functions of $m_t$.
\smallskip
We take this opportunity to reiterate that the effective Lagrangian
constructed here is to be used in the true sense of a Lagrangian.  In
particular, it may be used in a loop expansion.  All the basic properties
of the S-matrix, such as unitarity and analyticity are respected.  We also
want to comment on a very important issue, which is often misunderstood
and misapplied:  The operators $O_i$ in general have mass dimensions
higher than four.  They are termed 'unrenormalizable'.  However, in the
context of having its genesis from a renormalizable full theory, this is
truly a bad misnomer.  In fact, the reduction of a full theory into a
low energy effective theory guarantees that {\it $L_{eff}$ generates its own
renormalization program automatically} for these higher dimensional
operators, otherwise
there would be more free parameters than a renormalizable full theory could
accomodate.  To state this more explicitly, the only infinities in the
Green's functions constructed through $L_{eff}$ are those associated with
wave function, mass and coupling renormalizations.  All the other
intermediate infinities in a complete consistent calculation cancel.
The cancellation
occurs order by order in a loop expansion of $L_{eff}$ and makes
reference to the full theory only through $C_i$.  It then follows
that it makes no sense to deduce any physics from these spurious infinities,
as long as the underlying theory is renormalizable.  This point has only
recently been appreciated by some in phenomenological analysis [4].
\smallskip
In developing a loop expansion and simultaneously a derivative expansion,
there is a dimensional counting rule, which may be used to guide us to some
potentially interesting processes with the largest top effects.  Let
$$N=n_\partial +n_V +Max(m_t),$$
where
$$n_\partial=number \  of \  derivatives \  in \ O_i,$$
$$n_V=number \ of
  \ vector \ bosons \ in \ O_i,$$
and
$$Max(m_t)=maximum \ powers \ of \ m_t \ in \ C_i$$
Also, let
$$L=number \ of \ loops \ in \ a \ calculation, \ excluding
\ QCD \ corrections,$$
 and
$$n_f=number \ of  \ light \ quarks \ in \ O_i,$$
then [2]
$$N=2(L+1)-{n_f\over 2.} \eqno(1.3)$$
\smallskip
Thus, for a given loop order, once we specify the number of external light
fermions, we know what the maximum enhancement in power the top mass can give.
Particularly, to beat down the background, we are invited to consider flavor
changing neutral current processes $q_i \rightarrow q_j$, where $i (\neq j) =
b,s,d$.  Then top effects are produced by $q_i\rightarrow t \rightarrow
q_j$, which calls for the quantity $V_{ti}V_{tj}^*$.  This is a small
number, which may be compensated by large $m_t$ powers to make the
signal detectable.
\smallskip
Now, at the tree level, it is not difficult to show that there is no explicit
$m_t$ dependence.
At the one loop (L=1) level, there are certain cases which are known and
interesting.  Let us discuss those with dominant $m_t^2$ effects:
\smallskip
\noindent
(a) $n_f=n_V=n_\partial =0,$

\noindent
then, $Max(m_t)=4$.  This gives the modified Higgs potential, which in
turn places a loose bound on the Higgs mass for a given top
mass [5].
\smallskip
\noindent
(b) $n_f=0, \ n_\partial +n_V=2,$

\noindent
then, $Max(m_t)=2.$ This gives the celebrated $\rho (={\delta m_W^2
\over m_W^2}-{\delta m_Z^2 \over m_Z^2})$-parameter [6]. Also, we can
use it to find out the $m_t^2$ coefficient of $\sigma \rightarrow
W^+W^-$ [7], where $\sigma $ is the physical
Higgs.
\smallskip
\noindent
(c) $n_f=2, \ n_\partial +n_V=0.$

\noindent
This gives light quark mass renormalizations, where $Max(m_t)$=2 instead
of 3 from chiral argument.  Also, it gives us the $m_t^2$ coefficient of
$\sigma \rightarrow b\bar b$ [16], $b \bar s$, for example.
\smallskip
\noindent
(d) $n_f=2, \ n_\partial +n_V=1.$

\noindent
$Max(m_t)$ is 2. This gives rise to quark wave function
renormalizations and processes such as $b\rightarrow s+g$ [8], \ $s+\gamma$
[9].
In fact, after renormalization, the dependence becomes $O(m_t^0)$ due
to gauge invariance.  On the other hand, because of broken symmetry,
$Z \rightarrow b+ \bar s$ [10], \ $b+\bar b$ [11] still have dominant
$O(m_t^2)$ in amplitudes.
\smallskip
\noindent
(e) $n_f=4, \ n_\partial +n_V=0.$

\noindent
Here $Max(m_t)=2.$ Some processes of interest are $B-\bar B$-mixing [12]
and $\epsilon '$ in $K-\bar K$-mixing.
\smallskip
We have classified all the operators and obtained the effective
coupling coefficients to an accuracy of $O(m_t^0)$ for pure bosonic
processes, which cover (a), (b).  In this article, we want to deal
mainly with (c) and (d), where we shall again determine $C_i$ to
$O(m_t^0)$.  To be precise, we want to construct vertices with two
external light quarks and an arbitrary number of scalars, together
possibly with some vector bosons.  For the case at hand in general we
can actually have up to three derivatives in the operators $O_i$.
There will be then an enormous number of monomials, the coefficients
of which are to be determined.  We shall limit ourselves to those with
at most one derivative, so that some of the coefficients $C_i$ can
have $O(m_t^2)$ dominant dependence.  (We have in fact partially
evaluated the coefficients for vertices with two derivatives.  These
will be given in an appendix.)
\smallskip
Let us first consider the situation in which there are only the light
bottom quark and the heavy top quark, which form a doublet.  Later on
when we come to applications, we shall easily generalize the results to
include the other light quarks [13].

\smallskip
The plan of this article is as follows: In the next section we shall
recapitulate some of the notation and results obtained before.  Most
of the machinery we need for the derivation of $L_{eff}$ in this
sector has been developed in our previous publications.  One new piece
of tool is the bosonic propagators in external scalar fields, because
some of the internal lines are bosons.  We immediately go on to
construct these bosonic propagators.  The formal solutions of some of
the scalar propagators can be quite formidable, but we have been able
to cast them into a form such that derivative expansion can be
straightforwardly carried out by iteration.
\smallskip
In section 3, we outline how the effective vertices and their
coefficients are determined.  We shall describe how we convert a
generating Green's functional into an effective Lagrangian.  It is
done on each diagram with top internal line(s), which amounts to
performing algebraic rearrangement.  Although we do not want to dwell
upon it again, in short this procedure is an extension
of the Zimmermann oversubtraction scheme [14]. It
shows distinctly how effective vertices are developed and how an
automatic renormalization program is supplied in the process.  In
essence, {\it this is the foundation of the theory of effective
Lagrangians} , without which one has no mathematical justification for
its methodology and applicability.
\smallskip
In section 4, we use two examples $Z \rightarrow b \bar s$ and $Z
\rightarrow b \bar b$ to illustrate how easy and systematic it is to
reproduce known results.  As another example,
we evaluate the effective $\sigma b\bar{b}$ vertex with $O(m_t^2)$ dependence.
\smallskip
A brief conclusion will follow and several appendices will be supplied
to deal with some technical issues.

\noindent{\bf 2. BOSONIC PROPAGATORS IN EXTERNAL FIELDS}

In a previous publication [2], we constructed an effective Lagrangian which
incorporates all the virtual top quark effects contributing to radiative
corrections of
purely bosonic processes. In this special case, one needs to
consider only those diagrams composed  entirely of fermionic internal lines.
A technique was developed
to perform derivative expansion on the fermionic propagators as a part of
the general machinery. Here,
to construct an effective Lagrangian which contains two
external fermions, plus bosons as well, one has to compute diagrams
where there are also internal
bosons. It is then necessary to develop
a similar technique to deal with internal bosonic
propagators. To commence,
let us first recapitulate some results
from our previous paper. First of all, we write the standard model Lagrangian
as follows:
$${\it {L}}_{SM}={\it {L}}_{fermions}+{\it {L}}_{Higgs}
+{\it {L}}_{gauge \ fields}+{\it {L}}_{Yukawa}+{\it {L}}_{gauge \ fixing},
\eqno(2.1)$$
with
$${\it {L}}_{fermions}=-(\bar t \ \ \bar b)_L\gamma_{\mu} {1\over i} D^\mu
\pmatrix {t \cr b \cr }_L -\bar t_R \gamma _{\mu} {1\over i}D^\mu t_R
-\bar b_R\gamma_{\mu}{1\over i} D^\mu b_R,\eqno(2.2)$$
$${\it {L}}_{Higgs}=-(D^\mu \phi )^{\dagger }(D_{\mu} \phi )
-{\lambda \over 2}(\bar \phi \phi )^2-\mu ^2(\bar \phi \phi ),\eqno(2.3)$$
$${\it {L}}_{gauge \ fields}=-{1 \over 4}G^{a  \mu \nu}G^a_{\mu \nu}
-{1 \over 4}F^{\mu \nu}F_{\mu \nu},\eqno(2.4)$$
$${\it {L}}_{Yukawa}=-\Bigl[H(\bar t \ \ \bar b)_L \pmatrix {\phi ^0 \cr
\phi ^- \cr}t_R +h (\bar t \ \ \bar b)_L \pmatrix {-\phi ^+ \cr
\phi ^{0 \dagger } \cr}b_R +h.c. \Bigr],\eqno(2.5)$$
where
$$D_\mu \pmatrix {t \cr b \cr}_L = (\partial _\mu -ig {\tau ^a\over 2}A^a_\mu
-i{g'\over 2}Y_LB_\mu )\pmatrix {t \cr b \cr}_L,\eqno(2.6)$$
$$D_\mu t_R=(\partial _\mu -i{g'\over 2}Y_{t_R}B_\mu )t_R,\ \
D_\mu b_R=(\partial _\mu -i{g'\over 2}Y_{b_R}B_\mu )b_R,\eqno(2.7)$$
$$D_\mu \phi =(\partial _\mu -ig {\tau ^a \over 2}A^a_\mu -i{g' \over 2}
Y_\phi B_\mu )\phi ,\eqno(2.8)$$
$$G^a_{\mu \nu }=\partial _\mu A^a_\nu -\partial _\nu A^a_\mu +
g\epsilon ^{abc}A^b_\mu A^c_\nu , \ \ \ F_{\mu \nu}=\partial _\mu B_\nu
-\partial _\nu B_\mu ,\eqno(2.9)$$
The hypercharges are $Y_L= {1 \over 3}, \ Y_{t_R}={4 \over 3}, \ Y_{b_R}
=-{2 \over 3},$ and $Y_\phi =-1.$
We shall specify the gauge fixing terms in a moment.

The top quark acquires a mass $m_t={Hv\over \sqrt{2}}$ through the mechanism
of spontaneous symmetry breaking with $v$ being the symmetry breaking scale.
In the
limit $H\rightarrow \infty$, the $SU(2)\times U(1)$ gauge symmetry
is nonlinearly realized. Precisely speaking, the top quark field satisfies
the following constraints at the tree level [2]:
$$(\bar t \ \ \bar b)_L\pmatrix{\phi^0\cr \phi^-\cr}=0,\eqno(2.10)$$
and
$$t_R=0.\eqno(2.11)$$
These constraints are obtained by solving classical equations of motion
in the limit $H\rightarrow \infty$. Substituting Eqs. (2.10) and (2.11) into
$L_{SM}$, we obtain the tree level effective Lagrangian
$L_{n.l.}$:
$$\eqalign{{\it L}_{n.l.}=&-(\bar {\tilde t} \ \ \bar {\tilde b})_L
\gamma_{\mu} {1\over i}D^\mu
\pmatrix {{\tilde t} \cr {\tilde b} \cr }_L
-\bar {\tilde b}_R\gamma_{\mu} {1\over i}D^\mu \tilde b_R\cr
&-(D^{\mu} \tilde \phi)^{\dagger}(D_\mu \tilde \phi)-{\lambda \over 2}
(\bar {\tilde  \phi } \tilde \phi )^2
- \mu ^2(\bar {\tilde  \phi } \tilde \phi ) \cr
& - {1 \over 4 } \tilde G^{a  \mu \nu } \tilde G^a_{\mu \nu}
-{1\over 4} \tilde F^{\mu \nu} \tilde F_{\mu \nu} \cr
&-(h(\bar {\tilde b}_L\tilde \phi^{0 \dagger} -\bar {\tilde t}_L \tilde \phi
^+)
\tilde b_R + h.c.),  \cr}
\eqno(2.12)$$
where all quantities with tildes are solutions of nonlinear equations of
motion derived from $L_{n.l.}$. Note that ${\tilde t}_L=-({\tilde
\phi}^+/{\tilde \phi}^{0\dagger}){\tilde b}_L$.

To discuss one-loop corrections such that mainfest gauge invariance is
maintained, we perform a background field expansion
on the generating functional $W$ with background fields chosen to be solutions
of nonlinear equations of motion. First of all, the generating functional is
$$\eqalign{exp
\Bigl[iW[J,&K,J_\phi , \eta,\tilde A , \tilde B, \tilde \phi]\Bigr]
=\int dbd\bar bdtd\bar tdA_\mu dB_\mu d\phi d\bar {\phi}\Delta [A,B,\phi ,
\bar \phi; \tilde A,\tilde B,\tilde \phi , \bar {\tilde \phi }] \cr
& \cdot exp\Bigl[i\int dx(L_{SM}+\bar{\eta}_Rb_R+\bar{\eta}_Lb_L
+\bar{\phi}J_{\phi}+J_{\mu}^aA^{a \mu }+K_{\mu}B^{\mu})\Bigr],\cr }
\eqno(2.13)$$
where $\Delta $ is the Faddeev-Popov determinant corresponding to gauge
fixing terms given by
$${\it L}_{gauge \ fixing} =-{1 \over 2} G^{a\dagger}G^a-{1\over 2}G_B^\dagger
G_B ,\eqno(2.14)$$
with
$$G^a\equiv D^{\mu ab}(\tilde A)(A^b_{\mu}-\tilde A^b_{\mu})
+i{g\over 2}\bar {\tilde \phi}
\tau ^a (\phi -\tilde \phi)-i{g\over 2}(\bar \phi -\bar {\tilde \phi})
\tau ^a\tilde \phi, $$
and
$$G_B\equiv \partial ^\mu (B_{\mu} -\tilde B_{\mu} )-i{g'\over 2}
\bar {\tilde \phi}
(\phi - \tilde \phi )+i{g'\over 2}(\bar \phi - \bar {\tilde \phi})
\tilde \phi. \eqno(2.15)$$
The covariant derivative on the background field $D_\mu^{ab}(\tilde A)$
is defined as
$$D_\mu ^{ab}(\tilde A)= \partial _\mu \delta ^{ab}-g\epsilon ^{abc}
\tilde A_\mu ^c.\eqno(2.16)$$
We have also introduced extenal sources $\eta $'s, J's and K to induce
excitations.
Now we make a change of integration variables by shifting
$$b_L=B_L+\tilde b_L, \ \ \ b_R=B_R+\tilde b_R,$$
$$t_L=T_L+\tilde t_L, \ \ \ t_R=T_R$$
$$\phi=\Phi+\tilde \phi ,$$
$$A_\mu =A_\mu ^{q.f.}+\tilde A_\mu ,$$
and
$$B_\mu = B_\mu ^{q.f.}+\tilde B_\mu ,\eqno(2.17)$$
with $B_{L,R},T_{L,R},\Phi ,A_\mu ^{q.f.}$, and $B_\mu ^{q.f.}$ as quantum
fluctuations.
As a result, the generating functional $W$ becomes
$$\eqalign{exp\Bigl[iW\Bigr]& =exp\Bigl[i\int dx {\it {L}}_{n.l.} \Bigr] \cr
&\times \int dBd\bar BdTd\bar TdA_\mu ^{q.f.}dB_\mu ^{q.f.}d\Phi d\bar \Phi
\Delta exp\Bigl[i\int dx{\it {L}}'\Bigr],\cr }$$
where
$$\eqalign{{\it {L}}'=& \Bigl[(\bar B_L\tilde \phi ^-+\bar T_L\tilde \phi ^0)
\tilde t_R^{(1)}
+(\Phi ^0\bar {\tilde t}_L+\Phi ^-\bar {\tilde b}_L)\tilde t_R^{(1)}+h.c.
\Bigr] \cr
&+ terms \ \ in \ \  higher \ \ powers \ \ of \ \ quantum \ \  fluctuations,
 \cr }\eqno(2.18)$$
with
$${\tilde t}_R^{(1)}=(-\gamma_{\mu}{1\over i} [D^\mu \pmatrix {{\tilde t}_L \cr
{\tilde b}_L \cr }]^ \uparrow + h{\tilde \phi} ^+{\tilde b}_R)/{\tilde \phi^0}.
\eqno(2.19)$$
The up arrow means that the upper component of the term should be taken.
One can show that $t_R^{(1)}$ is not a singlet under $SU(2)$
transformation.  It is however a $SU(2)$ singlet
with respect to the $S$ matrix as we have shown in the previous paper.

Equipped with basic results above, we are ready to
discuss the behavior of various bosonic propagators in external background
fields. As we shall see, our calculational procedure requires us to
consider very few processes with external charged scalars.  Therefore,
only neutral-scalar background fields need to be included in constructing
propagators. We keep only the quadratic terms
 in Eq.(2.18) in deriving equations for
propagators.  The linear and other higher order fluctuations will be
treated as perturbations.
Let us begin our discussions by considering
the neutral scalar
propagators $<T(\Phi^0(x)\Phi^{0\dagger}(y))>$ and $<T(\Phi^{0\dagger}(x)
\Phi^{0\dagger}(y))>$. They satisfy certain equations of motion which
can be derived from the quadratic quantum fluctuations within $L^{\prime}$.
The relevant Lagrangian for deriving the equations of motion is given by
$$\eqalign{L^{\prime}_{n.quad.}=&-(\partial^{\mu}\Phi^{0\dagger})(\partial
_{\mu}\Phi^0)-\mu^2(\Phi^{0\dagger}\Phi^0)\cr
&+({g^2+g^{\prime 2}\over 8}-{\lambda\over 2})((\tilde \phi^{0\dagger})^2
(\Phi^0)^2+(\tilde \phi^{0})^2(\Phi^{0\dagger})^2)\cr
&-({g^2+g^{\prime 2}\over 4}
+2\lambda)\tilde{\phi^0}\tilde{\phi^{0\dagger}}\Phi^0\Phi^{0\dagger},\cr}
\eqno(2.20)$$
All quantities with tildes represent the unquantized background fields.
Gauge fixing terms give rise to contributions which are proportional to
$g^2+g^{\prime 2}$. Note that we do not use covariant derivatives here.  We
shall impose gauge invariance in our final results by minimal substitution,
together with adding terms with explicit field strength $F_{\mu \nu}$ and
$G_{\mu \nu}^a$ dependence.  The equation of motion for the field $\Phi^0$
follows from the action principle
$$\partial_{\mu}{\partial L^{\prime}_{n.quad.}\over \partial
(\partial_{\mu}\Phi^{0\dagger})}
={\partial L^{\prime}_{n.quad.}\over \partial \Phi^{0\dagger}}\eqno(2.21)$$
A similar equation holds for the conjugate field $\Phi^{0\dagger}$ by
replacing the field $\Phi^{0\dagger}$ in Eq. (2.21) with $\Phi^0$.
Now it is straightforward to derive equations of motion for propagators
of quantized neutral scalars in the background of their classical counterparts.
We obtain
$$\eqalign{&[\partial_x^2-\mu^2-(2\lambda+{g^2+g^{\prime 2}\over 4})
\tilde \phi^0 \tilde \phi^{0\dagger}]<T(\Phi^{0\dagger}(x)\Phi^{0\dagger}(y))>
\cr
&=(\lambda-{g^2+g^{\prime 2}\over 4})({\tilde \phi^{0\dagger}}(x))^2
<T(\Phi^0(x)\Phi^{0\dagger}(y))>,\cr}\eqno(2.22)$$
and
$$\eqalign{&[\partial_x^2-\mu^2-(2\lambda+{g^2+g^{\prime 2}\over 4})
\tilde \phi^0 \tilde \phi^{0\dagger}]<T(\Phi^0(x)\Phi^{0\dagger}(y))>
\cr
&=(\lambda-{g^2+g^{\prime 2}\over 4})({\tilde \phi^0}(x))^2
<T(\Phi^{0\dagger}(x)\Phi^{0\dagger}(y))>
-{1\over i}\delta^4 (x-y).\cr}\eqno(2.23)$$
To derive equations of motion for charged scalars, we begin with the following
Lagrangian:
$$\eqalign{L^{\prime}_{c.quad.}=&-(\partial^{\mu}\Phi^+)(\partial
_{\mu}\Phi^-)-\mu^2(\Phi^+\Phi^-)\cr
&-({g^2\over 2}
+\lambda)\tilde{\phi^0}\tilde{\phi^{0\dagger}}\Phi^+\Phi^-.\cr}\eqno(2.24)$$
$L^{\prime}_{c.quad.}$ is part of the quadratic quantum
fluctuations within $L^{\prime}$. Each term inside $L^{\prime}_{c.quad.}$
contains a pair of charged quantum fields as well as neutral background
fields. The equation of motion for $\Phi^+$ again follows from the action
principle
$$\partial_{\mu}{\partial L^{\prime}_{c.quad.}\over \partial
(\partial_{\mu}\Phi^-)}
={\partial L^{\prime}_{c.quad.}\over \partial \Phi^-}\eqno(2.25)$$
{}From Eq. (2.25), one can easily derive the equation of motion for the
propagator
\nextline
$<T(\Phi^+(x)\Phi^-(y))>$. It is
$$\eqalign{&[-\partial_x^2+\mu^2+(\lambda+{g^2\over 2})
\tilde \phi^0 \tilde \phi^{0\dagger}]<T(\Phi^+(x)\Phi^-(y))>
\cr
&={1\over i}\delta^4 (x-y).\cr}\eqno(2.26)$$

To complete the procedure, we derive
equations of motion for gauge field propagators. By
symmetry, it is sufficient to look at  $W$
boson propagators. First of all, the quadratic quantum fluctuations
involving the quantum fields $W^{\pm}$
and the classical neutral scalars read:
$$L^{\prime}_{w.quad.}=W^+_{\mu}g^{\mu\nu}(\partial^2
-{g^2\over 2}\tilde \phi^{0\dagger}\tilde \phi^0)W^-_{\nu}.\eqno(2.27)$$
The equation of motion for the field $W^+_{\alpha}$
can be derived via the action principle
$$\partial_{\beta}{\partial L^{\prime}_{w.quad.}\over \partial
(\partial_{\beta}W^-_{\alpha})}={\partial L^{\prime}_{w.quad.}
\over \partial W^-_{\alpha}},\eqno(2.28)$$
from which we obtain the equation of motion for
$<T(W^+_{\alpha}(x)
W^-_{\beta}(y))>$
$$\eqalign{&[-\partial_x^2+{g^2\over 2}
\tilde \phi^0 \tilde \phi^{0\dagger}]<T(W^+_{\alpha}(x)
W^-_{\beta}(y))>\cr
&={1\over i}g_{\alpha\beta}\delta^4 (x-y).\cr}\eqno(2.29)$$
Now Eqs. (2.22), (2.23), (2.26) and (2.29) can be solved by iteration
as will be elaborated below. It is easy to see that Eqs. (2.22) and (2.23)
are coupled equations which have to be solved simultaneously. As a first
step, we disentangle them as follows:
$$\eqalign{&[-\partial_x^2+\mu^2+(2\lambda+{g^2+g^{\prime 2}\over 4})
\tilde \phi^0 \tilde \phi^{0\dagger}]<T(\Phi^{0\dagger}(x)\Phi^{0\dagger}(y))>
\cr
&=\int d^4z(\lambda-{g^2+g^{\prime 2}\over 4})({\tilde \phi^{0\dagger}}(x))^2
[<x|{1\over -\partial^2+\mu^2+(2\lambda+{g^2+g^{\prime 2}\over 4})
\tilde \phi^0 \tilde \phi^{0\dagger}}|z>\cr
&\times (\lambda-{g^2+g^{\prime 2}\over 4})({\tilde \phi^0}(z))^2
<T(\Phi^{0\dagger}(z)\Phi^{0\dagger}(y))>\cr
&+{1\over i}<x|{1\over -\partial^2+\mu^2+(2\lambda+{g^2+g^{\prime 2}\over 4})
\tilde \phi^0 \tilde \phi^{0\dagger}}|z>\delta^4(z-y)],\cr}\eqno(2.30)$$
and
$$\eqalign{&[-\partial_x^2+\mu^2+(2\lambda+{g^2+g^{\prime 2}\over 4})
\tilde \phi^0 \tilde \phi^{0\dagger}]<T(\Phi^0(x)\Phi^{0\dagger}(y))>
\cr
&=\int d^4z(\lambda-{g^2+g^{\prime 2}\over 4})({\tilde \phi^0}(x))^2
[<x|{1\over -\partial^2+\mu^2+(2\lambda+{g^2+g^{\prime 2}\over 4})
\tilde \phi^0 \tilde \phi^{0\dagger}}|z>\cr
&\times (\lambda-{g^2+g^{\prime 2}\over 4})({\tilde \phi^{0\dagger}}(z))^2
<T(\Phi^0(z)\Phi^{0\dagger}(y))>]
+{1\over i}\delta^4(z-y).\cr}\eqno(2.31)$$
Note that, as they stand, these are integral-differential equations, which
are not
particularly useful for derivative expansion.  We need to make certain
rearrangement to make the non-locality controllable by iteration.  For
convenience, we introduce the operator $\Delta(\Phi^{0\dagger}
\Phi^{0\dagger})$ which is defined as
$$<T(\Phi^{0\dagger}(x)\Phi^{0\dagger}(y))>
\equiv <x|\Delta(\Phi^{0\dagger}\Phi^{0\dagger})|y>.\eqno(2.32)$$
A similar definition holds for the operator
$\Delta(\Phi^0\Phi^{0\dagger})$. We shall convert Eqs. (2.30) and
(2.31) into operator equations of $\Delta(\Phi^{0\dagger}
\Phi^{0\dagger})$ and $\Delta(\Phi^0\Phi^{0\dagger})$. After some
algebraic manipulation, we arrive at
$$\eqalign{\Delta(\Phi^{0\dagger}\Phi^{0\dagger})=&{1\over 2}
(O_1+O_2)(\lambda-{g^2+g^{\prime 2}\over 4})u\cr
&\times [{\tilde \phi^{0\dagger}\over \tilde \phi^0},O_3]{\tilde \phi^0\over
\tilde \phi^{0\dagger}}(\lambda-{g^2+g^{\prime 2}\over 4})u
\Delta(\Phi^{0\dagger}\Phi^{0\dagger})\cr
&-{1\over 2}(O_1+O_2)(\lambda-{g^2+g^{\prime 2}\over 4})u
{\tilde \phi^{0\dagger}\over \tilde \phi^0}{1\over i}O_3,\cr}\eqno(2.33)$$
and
$$\eqalign{\Delta(\Phi^0\Phi^{0\dagger})=&{1\over 2}
(O_1+O_2)(\lambda-{g^2+g^{\prime 2}\over 4})u\cr
&\times [{\tilde \phi^0\over \tilde \phi^{0\dagger}},O_3]{\tilde
\phi^{0\dagger}
   \over
\tilde \phi^0}(\lambda-{g^2+g^{\prime 2}\over 4})u
\Delta(\Phi^0\Phi^{0\dagger})\cr
&+{1\over 2i}(O_1+O_2),\cr}\eqno(2.34)$$
where $u \equiv \tilde \phi^0 \tilde \phi^{0\dagger}$; also
$$O_1 \equiv {1\over -\partial^2+\mu^2+3\lambda u},\eqno(2.35a)$$
$$O_2 \equiv {1\over -\partial^2+\mu^2+(\lambda+{g^2+g^{\prime 2}\over 2})u},
\eqno(2.35b)$$
and
$$O_3 \equiv {1\over -\partial^2+\mu^2+(2\lambda+{g^2+g^{\prime 2}\over 4})u}
\eqno(2.35c)$$
Note the occurence of commutators in Eqs. (2.33) and (2.34). They generate
at least one more derivative.  The equations are now
suitable for recursive iteration to any
desirable order in derivative expansions.

The equations of motion for propagators of charged scalars and $W$ bosons
are much easier to solve. From Eqs. (2.26) and (2.29), we have
$$\Delta(\Phi^+\Phi^-)={1\over i}{1\over -\partial^2+\mu^2
+(\lambda+{g^2\over 2})u},\eqno(2.36)$$
and
$$\Delta(W^+_{\alpha}W^-_{\beta})={1\over i}g_{\alpha\beta}
{1\over -\partial^2+{g^2\over 2}u},
\eqno(2.37)$$
where $\Delta(\Phi^+\Phi^-)$ and $\Delta(W^+_{\alpha}W^-_{\beta})$ are
again related to the their corresponding propagator functions
as illustrated before. Now one can likewise expand the above equations
to any given order in derivatives.

\noindent{\bf 3. TWO-FERMION EFFECTIVE LAGRANGIAN}

In the last section, we derived equations of motion for
various bosonic propagators, the solutions of which can be expanded
in derivatives. The expansion scale is $\sim <\bar \phi \phi >={v^2\over 2}$.
We are now ready to construct an effective Lagrangian
containing two light external fermions, which in the current context
refer to the bottom quarks, and arbitrary numbers of
external scalar bosons.
According to a power counting argument, for a $L$ loop diagram with $n_F$
light external fermions, the order
of derivative expansion one needs to perform for achieving an accuracy
such that the neglected terms are suppressed by at least $O({1\over m_t})$
is given by [2]

$$N=2(L+1)-{1\over 2}n_F.\eqno(3.1)$$
In view of this, one should in principle compute the two-fermion
effective Lagrangian up to three derivatives.
This also implies that the coefficients of the potential terms could be
$O(m_t^3)$ in their $m_t$ dependence.
In actuality, it is easy to show that
the coefficients should always be
even functions of $m_t$. If we just want to focus on $O(m_t^2)$  contributions,
it is sufficient
for us to compute the potential and one-derivative terms of the
effective Lagrangian. The coefficients of the
two-derivative and three-derivative terms  depend on  $m_t$ only
logarithmically, which will be addressed later on.
In the following, we list all the possible
gauge invariant two-fermion operators carrying zero and one derivative:

\noindent{\bf{Potential terms}}
$$A\bar{\psi}_L\phi^cb_R+h.c.,\eqno(3.2)$$
where
$$\bar{\psi}_L=(\bar{t}_L^0 \ \ b_L); \ \phi^c=\pmatrix{-\phi^+ \cr
\phi^{0\dagger} \cr}.\eqno(3.3)$$
We have already dropped all the tildes associated with classical fields.
This will not cause any confusion because we can easily distinguish them
from quantum fields, which are denoted by upper case letters.
The quantity $t_L^0$ in Eq. (3.3) was
originally denoted by $\tilde t_L$, which is related to $\tilde b_L$ and scalar
fields
through the nonlinear constraint given by Eq. (2.10).

\noindent{\bf{One-derivative terms}}
$$\eqalign{&a_1(\bar{\psi}_L{1\over i}\gamma \cdot \overrightarrow D\psi_L
-\bar{\psi}_L{1\over i}\gamma \cdot \overleftarrow D^{\dagger}\psi_L),\cr
&a_2(\bar{\psi}_L{1\over i}\gamma \cdot \overrightarrow D\psi_L
-\bar{\psi}_L{1\over i}\gamma \cdot \overleftarrow D^{\dagger}\psi_L)
ln({\bar{\phi}\phi\over \kappa^2}),\cr
&a_3\bar{\psi}_L{1\over i}\gamma_{\mu}\psi_L[(D^{\mu}\phi)^{\dagger}\phi
-\bar{\phi}(D^{\mu}\phi)]/\bar{\phi}\phi,\cr
&a_4\bar{\psi}_L{1\over i}\gamma_{\mu}\psi_L[(D^{\mu}\phi)^{\dagger}\phi
-\bar{\phi}(D^{\mu}\phi)]ln({\bar{\phi}\phi\over \kappa^2})
/\bar{\phi}\phi,\cr
&a_5\mu^2(\bar{\psi}_L{1\over i}\gamma \cdot \overrightarrow D\psi_L
-\bar{\psi}_L{1\over i}\gamma \cdot \overleftarrow
D^{\dagger}\psi_L)/\bar{\phi}
   \phi,\cr
&a_6\mu^2(\bar{\psi}_L{1\over i}\gamma \cdot \overrightarrow D\psi_L
-\bar{\psi}_L{1\over i}\gamma \cdot \overleftarrow D^{\dagger}\psi_L)
ln({\bar{\phi}\phi\over \kappa^2})/\bar{\phi}\phi,\cr
&a_7\mu^2\bar{\psi}_L{1\over i}\gamma_{\mu}\psi_L[(D^{\mu}\phi)^{\dagger}\phi
-\bar{\phi}(D^{\mu}\phi)]/(\bar{\phi}\phi)^2,\cr
&a_8\mu^2\bar{\psi}_L{1\over i}\gamma_{\mu}\psi_L[(D^{\mu}\phi)^{\dagger}\phi
-\bar{\phi}(D^{\mu}\phi)]ln({\bar{\phi}\phi\over \kappa^2})
/(\bar{\phi}\phi)^2,\cr
&b_1(\bar{b}_R{1\over i}\gamma \cdot \overrightarrow Db_R
-\bar{b}_R{1\over i}\gamma \cdot \overleftarrow D^{\dagger}b_R)
,\cr
&b_2(\bar{b}_R{1\over i}\gamma \cdot \overrightarrow Db_R
-\bar{b}_R{1\over i}\gamma \cdot \overleftarrow D^{\dagger}b_R)
ln({\bar{\phi}\phi\over \kappa^2}),\cr
&b_3\bar{b}_R{1\over i}\gamma_{\mu}b_R[(D^{\mu}\phi)^{\dagger}\phi
-\bar{\phi}(D^{\mu}\phi)]/\bar{\phi}\phi,\cr
&b_4\bar{b}_R{1\over i}\gamma_{\mu}b_R[(D^{\mu}\phi)^{\dagger}\phi
-\bar{\phi}(D^{\mu}\phi)]ln({\bar{\phi}\phi\over \kappa^2})
/\bar{\phi}\phi,\cr
&c_1(\bar{t}_R^{(1)}{1\over i}\gamma \cdot \overrightarrow Dt_R^{(1)}
-\bar{t}_R^{(1)}{1\over i}\gamma \cdot \overleftarrow D^{\dagger}t_R^{(1)})
,\cr
&c_2\bar{t}_R^{(1)}{1\over i}\gamma \cdot \overrightarrow Dt_R^{(1)}
-\bar{t}_R^{(1)}{1\over i}\gamma \cdot \overleftarrow D^{\dagger}t_R^{(1)})
ln({\bar{\phi}\phi\over \kappa^2}),\cr
&c_3\bar{t}_R^{(1)}{1\over i}\gamma_{\mu}t_R^{(1)}[(D^{\mu}\phi)^{\dagger}\phi
-\bar{\phi}(D^{\mu}\phi)]/\bar{\phi}\phi,\cr
&c_4\bar{t}_R^{(1)}{1\over i}\gamma_{\mu}t_R^{(1)}[(D^{\mu}\phi)^{\dagger}\phi
-\bar{\phi}(D^{\mu}\phi)]ln({\bar{\phi}\phi\over \kappa^2})
/\bar{\phi}\phi.\cr}\eqno(3.4)$$
We have used $\kappa$ to denote the renormalization scale. Every operator
here is $CP$ invariant because we have not included $KM$ mixing in the present
discussion. Also we want to remind the reader again that
$t_R^{(1)}$ is not a $SU(2)$ singlet unless we are
considering the $S$ matrix.

As demonstrated in our previous paper, the effective
Lagrangian constructed here will incorporate virtual top quark effects in every
one-light-particle irreducible diagram with external momenta less than the
top quark mass. We also pointed out that the determination of the one loop
effective Lagrangian is a two-step process. First of all, we compute the
one-loop effective action $\Gamma_{1LPI}^{1loop}$
with the derivative expansion technique. There it is assumed that
the external momenta for a given diagram are much less than the mass of
the top quark as well as the masses of the other internal particles.
We obtain a local effective
action which is written as
$$\Gamma_{1LPI}^{1loop}=\int d^4x \Omega_{1LPI}^{1loop}.\eqno(3.5)$$
It also follows that the applicability of $\Gamma_{1LPI}^{1loop}$
($\Omega_{1LPI}^{1loop}$) is
limited to the case where the external momenta are less than the
mass of any internal particle. This result comes short
of our expectations because
we would like to construct an effective Lagrangian applicable to any process
as long as its external momenta are less than $m_t$.
To achieve this goal, we
subtract from $\Omega_{1LPI}^{1loop}$
the one loop contributions $\Omega_{ind.}^{1loop}$
induced by the tree level effective
Lagrangian $L_{n.l.}$ up to the same order in derivatives. The result will
be the one loop effective Lagrangian we are after. Explicitly, we have
$$L_{eff}^{1loop}=\Omega_{1LPI}^{1loop}-\Omega_{ind.}^{1loop}.\eqno(3.6)$$
Now all the leading
heavy top effects to the one loop order are given by $L_{n.l.}$ and
$L_{eff}^{1loop}$. Since $L_{n.l.}$ actually gives rise to leading top
effects for tree level processes, we shall use the notation $L_{eff}^{tree}$
to replace it from now on.
To repeat then, to extract contributions from $L_{eff}^{1loop}$,
we simply read them off from the operators in Eqs (3.2) and (3.4).
Additional one-loop contributions are coming from $L_{eff}^{tree}$, where we
use it to construct loop diagrams and perform exact calculations.
S-matrix analyticity, such as threshold cuts, will then be respected.
However, if we are only after leading $m_t$ effects,
we can ignore this last piece of contributions because they do not contain
any $m_t$ dependence as one can readily see from Eq. (2.12).

For computing coefficients of the two-fermion effective Lagrangian,
let us expand the standard
model Lagrangian around the solutions of non-linear equations of motion.
At the one loop order, these coefficients receive contributions from
linear, quadratic, and cubic quantum fluctuations. The linear
fluctuations arises because we do not expand the Lagrangian around
exact solutions of classical equations of motion. The
cubic quantum fluctuations are needed because we also need to include
contributions from diagrams which are one particle reducible with respect to
internal
top quark lines. Such diagrams necessarily involve both linear and cubic
quantum fluctuations. In the following, we shall give only a part of the
entire Lagrangian and omit terms which are irrelevant
to our discussion. The linear, quadratic,
and cubic quantum fluctuations are given by
$$L^1=(\bar{B}_L\phi^- +\bar{T}_L\phi^0)t_R^{(1)}
+(\Phi^0\bar{t}_L^{(0)}+\Phi^-\bar{b}_L)t_R^{(1)} +h.c.,\eqno(3.7)$$
$$\eqalign{L^2=&-(\bar{T} \ \ \bar{B})_L{1\over i}\gamma \cdot \partial
\pmatrix {T\cr B\cr}_L -\bar{T}_R{1\over i}\gamma \cdot \partial T_R
-\bar{B}_R{1\over i}\gamma \cdot \partial B_R \cr
&-H(\bar{T} \ \ \bar{B})_L \pmatrix {\phi^0 \cr \phi^-\cr}T_R
-H(\bar{t}^{(0)} \ \ \bar{b})_L \pmatrix {\Phi^0 \cr \Phi^-\cr}T_R\cr
&-h(\bar{T} \ \ \bar{B})_L \pmatrix {-\phi^+ \cr \phi^{0\dagger}\cr}B_R
-h(\bar{t}^{(0)} \ \ \bar{b})_L \pmatrix {-\Phi^+ \cr \Phi^{0\dagger}\cr}B_R\cr
&-h(\bar{T} \ \ \bar{B})_L \pmatrix {-\Phi^+ \cr \Phi^{0\dagger}\cr}b_R
+h.c.\cr
&-{1\over 2}[D^{\mu ab}(A)A_{\mu}^{qf \ b}+{ig\over 2}
(\bar{\phi}\tau^a\Phi-\bar{\Phi}\tau^a\phi)]^{\dagger}\cr
&\cdot [D^{\nu ac}(A)A_{\nu}^{qf \ c}+{ig\over 2}
(\bar{\phi}\tau^a\Phi-\bar{\Phi}\tau^a\phi)]\cr
&-{1\over 2}[\partial^{\mu}B_{\mu}^{qf}-{ig^{\prime}\over 2}
(\bar{\phi}\Phi-\bar{\Phi}\phi)]^{\dagger}\cr
&\cdot[\partial^{\nu}B_{\nu}^{qf}-{ig^{\prime}\over 2}
(\bar{\phi}\Phi-\bar{\Phi}\phi)]+ \cdots\cr}\eqno(3.8)$$
$$L^3=-H(\bar{T} \ \ \bar{B})_L\pmatrix {\Phi^0\cr \Phi^-\cr}T_R
-h(\bar{T} \ \ \bar{B})_L\pmatrix{-\Phi^+\cr \Phi^{0\dagger}\cr}B_R
+h.c.+\cdots ,\eqno(3.9)$$
where as said we have dropped the tildes for background fields.
{}From $L^2$, one can derive equations of motion for
fermionic and bosonic propagators, as we already did.

To determine
potential terms, we choose to compute the $\bar{b}_Lb_R$ self energy.
In order to match the result of such a direct computation with the one
given by the effective Lagrangian $L_{eff}^{1loop}$,
one has to subtract from
the full $\bar{b}_Lb_R$ self energy a contribution induced by the tree level
effective Lagrangian $L_{eff}^{tree}$. This procedure was elucidated earlier
in Eq. (3.6) and is now further illustrated in
diagrams in Fig. 1 for this special case.
As depicted there, the diagram arising from
$L_{eff}^{tree}$ contains a four-point $\bar{b}_Lb_R\phi^+\phi^-$
vertex. This vertex is actually generated by shrinking the internal top quark
propagator $<T(T_R \ \bar{T}_L)>$ from the original $\bar{b}_Lb_R$
self energy diagram.
In fact, the process of
shrinking the propagator of a heavy particle is equivalent to performing
a momentum expansion (or derivative expansion in the configuration space)
on it. This expansion is terminated at the order of one external
momentum( or one derivative in the configuration space)
since all the higher order terms
are suppressed by $O({1\over m_t})$. The operation actually provides an
alternative method for computing $\Omega_{ind.}^{1loop}$. Instead of
computing one-loop diagrams
by using $L_{eff}^{tree}$, which is an approach we took in the
previous paper,
we may as well start with
the original diagram given by the full theory and perform a derivative
expansion on
each heavy top propagator where two-derivative and higher order terms
are neglected. As a result, each top quark propagator within the one-loop
diagram is shrunken and replaced with a corresponding effective vertex.
Such effective vertices are precisely those given by the effective
tree level Lagrangian $L_{eff}^{tree}$. This alternative approach has the
advantage of maintaining one to one correspondence between two sets
of diagrams which contribute to $\Omega_{1LPI}^{1loop}$ and
$\Omega_{ind.}^{1loop}$ respectively.
As shown in Eq. (3.6), we obtain $L_{eff}^{1loop}$
by computing the difference between two diagrams which have the same origin
, and all the terms with infrared
cutoff will cancel out as a result of such a subtraction.
Before we proceed
to compute the diagram in Fig. 1, let us first provide rules for shrinking
heavy top propagators within those one-loop diagrams which, after the
shrinking, will contribute to $\Omega_{ind.}^{1loop}$. For convenience, we
define an operator $\tau$ to denote such a shrinking. The rules for shrinking
various top quark propagators with different helicity configurations are listed
as follows:
$$\tau<T(T_L(x)\bar{T}_L(y))>=0,\eqno(3.10a)$$
$$\tau<T(T_R(x)\bar{T}_L(y))>={-i\delta^4(x-y)\over H\phi^0(x)}R,
\eqno(3.10b)$$
$$\tau<T(T_L(x)\bar{T}_R(y))>={-i\delta^4(x-y)
\over H\phi^{0\dagger}(x)}L,
\eqno(3.10c)$$
and
$$\eqalign{\tau<T(T_R(x)\bar{T}_R(y))>=&{\gamma^{\mu}\phi^{0\dagger}(y)
(\partial_{\mu}^y \phi^0(y))\over H^2(\phi^{0\dagger}\phi^0)^2}
\delta^4(x-y)L\cr
&+{\gamma^{\mu}\over H^2\phi^{0\dagger}\phi^0}\partial_{\mu}^x
\delta^4(x-y)L.\cr}\eqno(3.10d)$$
Remarks about these equations are in order. First of all,
every top quark propagator appearing above is in the background of
neutral scalar fields. Equations of motion for such propagators were
given by Eqs. (5.1) and (5.2) in our previous paper [15]. They can be
expanded in derivatives as we demonstrated in Eqs. (I-5.3) and (I-5.4).
In particular, we gave the solution for the propagator
$<T(T_R \ \bar{T}_R)>$ to four derivatives as shown in  Eq. (I-5.12). Secondly,
the action of the operator $\tau$
terminates at a certain number of derivatives. For the propagator
$<T(T_R \ \bar{T}_R)>$, we stop at potential and one-derivative terms.
For
$<T(T_R \ \bar{T}_L)>$ and its hermitian conjugate, only potential terms are
kept. Finally $\tau<T(T_L \ \bar{T}_L)>$ is set to zero. Such prescriptions
can be easily understood by looking at the tree level $b$-$\phi$ scattering
process mediated by internal top quark propagators. As shown in Fig. 2, the
diagram which contains the
propagator
$<T(T_L \ \bar{T}_L)>$ has two vertices all proportional to $m_b$ rather than
$m_t$. This indicates that even the leading term in the propagator
$<T(T_L \ \bar{T}_L)>$ would still render the entire $b$-$\phi$ scattering
amplitude of the order ${1\over m_t}({1\over H})$. This precisely accounts for
Eq. (3.10a).
For the diagram containing the propagator $<T(T_R \ \bar{T}_R)>$, the
two vertices are all proportional to $m_t$. Therefore one needs to
keep $O({1\over H^2})$ terms as shown in Eq. (3.10d). Further discussions
on this issue will be given in appendix A.

After the discussion of propagator shrinking, we now return to Fig. 1, which
lists diagrams to be computed for determining the potential term of the
effective Lagrangian. In order to have a top quark
appearing as an internal line,
the boson accompanying it in the loop has to be charged.
It could either be
$W^{\pm}$ or $\Phi^{\pm}$. The former is however ruled out because
$W^{\pm}$ only interacts with left-handed fermions. The first diagram
is derived from the Lagrangian in Eq. (3.8). Let us denote its amplitude
by $I_a$, which reads in configuration space as:
$$\eqalign{I_a=&-i^2Hh\int d^4x \int d^4y <T(\Phi^+(y)\Phi^-(x))>
\bar{b}_L(x)\cr
&\cdot <T(T_R(x)\bar{T}_L(y))>b_R(y).\cr}\eqno(3.11)$$
Following the method introduced in Eq. (I-6.11) of our previous paper, we
transform $I_a$ into
$$\eqalign{I_a=&-i^2Hh\int d^4y\int {d^4k\over (2\pi)^4}<\Phi^+\Phi^-(y-i{
\overleftarrow \partial\over \partial k})>[\bar{b}_L(y+i{
\overrightarrow \partial\over \partial k})\cr
&\cdot <T_R\bar{T}_L(y+i{
\overrightarrow \partial\over \partial k})>b_R(y)].\cr}\eqno(3.12)$$
The notations here are those in our previous paper. For the
second diagram, let us denote its amplitude by $I_b$. Now $I_b$ can be obtained
from $I_a$ by replacing the top quark propagator $<T(T_R(x)\bar{T}_L(y))>$
there with $\tau<T(T_R(x)\bar{T}_L(y))>$, given in Eq. (3.10b). Integrating
out the delta function $\delta^4(x-y)$ gives
$$I_b=-ih\int d^4y <T(\Phi^+(y)\Phi^-(y))>{\bar{b}_L(y)b_R(y)\over \phi^0}.
\eqno(3.13)$$
It is easy to check that an exactly identical $I_b$ can be obtained directly
from the tree effective Lagrangian $L_{eff}^{tree}$.
The calculation of $I_a-I_b$ is rather simple, the result of which is used to
fix the coefficient $A$ of the potential term:
$$\eqalign{A=&{1\over 16 \pi^2}h(H^2+\lambda+{g^2\over 2}
+{\mu^2\over \bar{\phi}\phi})\cr
&\cdot ({-1\over \epsilon}+\gamma_E+ln(\pi)-1+ln({H^2\bar{\phi}\phi\over
\kappa^
   2}))
.\cr}\eqno(3.14)$$

For computing one derivative terms, we choose processes which do not
flip the helicity of the fermions.
For determining coefficients $a_1$ through $a_8$,
the simplest choice is $\bar{b}_L-
b_L$ self energy diagrams. In Fig. 3, we list all the contributing diagrams.
In this process, a diagram with internal $W^{\pm}$ boson is allowed. Note
also that there is no subtraction needed for this diagram because  the internal
fermion propagator is $<T(T_L \ \bar{T}_L)>$ and $\tau<T(T_L \ \bar{T}_L)>=0$
according to Eq. (3.10a). Now the calculation is rather straightforward
and we give the final results:
$$\eqalign{&a_1={1\over 16 \pi^2}(\lambda+{1\over 2}H^2+g^2)({-1\over {2
\epsilon ^{\prime}}}+{1\over 2}ln(H^2)-{3\over 4}),\cr
&a_2={1\over 16 \pi^2}[{1\over 2}(\lambda+{1\over 2}H^2+g^2)],\cr
&a_3={1\over 16 \pi^2}[(\lambda+{1\over 2 }H^2)-(\lambda+{1\over 2}g^2)
({-1\over {2\epsilon^{\prime}}}+{1\over 2}ln(H^2))],\cr
&a_4={1\over 16 \pi^2}[-{1\over 2}(\lambda+{g^2\over 2})],\cr
&a_5={1\over 16 \pi^2}[{-1\over {2\epsilon^{\prime}}}-{3\over 4}
+{1\over 2}ln(H^2)],\cr
&a_6={1\over 16 \pi^2}({1\over 2}),\cr
&a_7={1\over 16 \pi^2}({1\over {2\epsilon^{\prime}}}+1-{1\over 2}ln(H^2)),\cr
&a_8={1\over 16 \pi^2}(-{1\over 2}),\cr}\eqno(3.15)$$
where ${1\over \epsilon^{\prime}}\equiv {1\over \epsilon}-(\gamma_E
+ln(\pi))$ and $\epsilon=2-{n\over 2}$.

Similarly, we compute the $\bar{b}_R-b_R$ self energy diagram to determine
coefficients $b_1$ through $b_4$. Only one diagram contributes as shown
in Fig. 4. Once more, no subtraction is needed due to the same reason
given before. We find
$$\eqalign{&b_1={1\over 16 \pi^2}[h^2({-1\over 4\epsilon^{\prime}}-{3\over 8}
+{1\over 4}ln(H^2))],\cr
&b_2={1\over 16 \pi^2}({1\over 4}h^2),\cr
&b_3={1\over 16 \pi^2}(-{1\over 2}h^2),\cr
&b_4={1\over 16 \pi^2}(0).\cr}\eqno(3.16)$$

The computation of $c_1$ through $c_4$ is more involved. This is because
the structure $t_R^{(1)}$ inevitably involves charged scalar $\phi^{\pm}$. We
thus need to look at  processes with two right-handed
fermions and a pair of charged scalars in external lines. To save effort,
we first perform a power counting argument to eliminate diagrams which
approach
to zero as $H\rightarrow \infty$. The remaining diagrams are listed in Fig. 5.
The diagrams in Fig. 5a and 5b can be handled in a standard way.  A new
feature here is the
appearance of one particle reducible diagrams with respect to
internal top quark lines. We briefly describe a procedure for
performing derivative expansion on reducible diagrams, which will
be applied to calculate diagrams shown in Fig. 5c-5e.
As an illustration, we
write down expressions for $I_c$ and $I_e$, which are amplitudes for
the diagrams in Fig. 5c and 5e, respectively. The expression for $I_d$ is
similar to that for $I_c$.  Thus, we have
$$\eqalign{I_c=&-H^2h^2\int d^4xd^4yd^4zd^4w\bar{b_R}(x)
<T(B_L(x)\bar{B}_L(y))>\phi^-(y)\cr
&\cdot <T(T_R(y)\bar{T}_R(z))>
<T(\Phi^{0\dagger}(z)\Phi^0(x))><T(T_L(z)\bar{T}_L(w))>\cr
&\cdot \phi^+(w)b_R(w.)\cr}\eqno(3.17)$$
Let us define
$$G(z)\equiv \int d^4w<T(T_L(z)\bar{T}_L(w))>\phi ^+(w)b_R(w).\eqno(3.18)$$
Then, taking into account the anticommuting property of fermions and
specifying the Dirac indices for them properly when we alter their
ordering (which are understood), we write

$$\eqalign{I_c=&H^2h^2\int d^4xd^4yd^4z<T(B_L(x)\bar{B}_L(y))>\phi^-(y)\cr
&\cdot <T(T_R(y)\bar{T}_R(z))>G(z)
<T(\Phi^{0\dagger}(z)\Phi^0(x))>\bar b_R(x)\cr
&=H^2h^2\int d^4x \int {d^4k\over {2\pi ^4}}<B_L\bar{B}_L\phi^-T_R\bar {T}_R
(x-i\overleftarrow {\partial \over {\partial k}})>\cr
&\cdot <G\Phi^{0 \dagger }\Phi^0(x+i\overrightarrow {\partial \over
{\partial k}})>\bar {b}_R(x).\cr}\eqno(3.19)$$
In the above, we have used momentum representation.  We should take
note that both strings of operators are functions of
momentum $k$ and position $i{\partial \over {\partial k}}$.  We have not
displayed the momentum dependence explicitly, however.  In the same
representation, we can easily derive
$$G(z)=<T_L\bar {T}_L\phi^+b_R(z-i\overleftarrow {\partial \over
{\partial p}})>|_{p=0}.\eqno(3.20)$$
Equivalently, we can use coordiante representation, in which
$$\eqalign {G(z)=&\int d^4w<z|T_L\bar{T}_L\phi^+b_R|w>\cr
&=\int d^4w<z|T_L\bar{T}_L\phi^+b_R(z_{op},p_{op})|w>,\cr}\eqno(3.21)$$
where we have explicitly displayed both momentum and coordinate
dependence.  Then,
$$\eqalign{G(z)=&<T_L\bar{T}_L\phi^+b_R(z,-i\overrightarrow {\partial
\over {\partial z}})>\cr
&=<<T_L\bar{T}_L\phi^+b_R(-i\overrightarrow {\partial
\over {\partial z}})>>,\cr}\eqno(3.22)$$
where we have used double pointed brackets to denote that coordinate
representation has been used and that the dependence on z has been understood.
This leads to
$$\eqalign {I_c=&H^2h^2\int d^4x \int {d^4k\over (2\pi)^4}<B_L\bar{B}_L
\phi^-T_R\bar{T}_R
(x-i{\overleftarrow{\partial}\over \partial k})>\cr
&\cdot [<<T_L\bar{T}_L(-i{
\overrightarrow\partial  \over {\partial z}})>>\phi^+(z)b_R(z)]|_{z=x+
i{\overrightarrow \partial \over {\partial k}}} \cr
& \cdot <\Phi^{0\dagger}\Phi^0(x+i\overrightarrow{\partial \over
{\partial k}})> \bar{b}_R(x).\cr}\eqno(3.23)$$
Similarly,
$$\eqalign{I_e=&H^2h^2\int d^4xd^4yd^4zd^4w\bar{b}_R(x)\phi^-(x)
<T(T_L(x)\bar{T}_R(y))>\cr
&\cdot <T(\Phi^{0\dagger}(y)\Phi^0(z))>
<T(T_L(y)\bar{T}_L(z))><T(T_R(z)\bar{T}_L(w))>\cr
&\cdot \phi^+(w)b_R(w)\cr
&=-H^2h^2\int d^4x\int {d^4k\over (2\pi)^4}<T_L\bar{T}_L(x-i
{\overleftarrow \partial \over {\partial k}})>\cr
&\cdot [<<T_R\bar{T}_L(-i
{\overrightarrow \partial \over {\partial z}})>>\phi^+(z)b_R(z)]|_{z=x+
i{\overrightarrow \partial \over {\partial k}}}
<\Phi^0\Phi^{0\dagger}(x+i{\overrightarrow \partial \over {\partial k}})>\cr
&\cdot [\bar{b}_R(x)\phi^-(x)<<T_L\bar{T}_R(i
{\overleftarrow \partial \over {\partial x}})>>].\cr}\eqno(3.24)$$
Again, we have altered the ordering of some of the fermion operators
in Eq.(3.24). To determine the coefficients
$c$'s, we sum over all the amplitudes for the process $b_R\phi^-\rightarrow
b_R\phi^-$
and equate the result with that given by
the effective Lagrangian. In fact, this particular process also receives
contributions from operators with coefficients $b_1$ to $b_4$. It is worth
noting that diagrams 5a and 5b will be entirely contributed by such
operators. This is not surprising because diagrams 5a and 5b are composed
from quadratic quantum fluctuations $L^2$ alone,
and therefore have nothing to do with the structure $t_R^{(1)}$.
The rest of the diagrams will simply involve operators with coefficients
$c_1$ to $c_4$. We then arrive at
$$\eqalign{&c_1={1\over 16 \pi^2}({-5\over 4\epsilon^{\prime}}-{11\over 8}
+{5\over 4}ln(H^2)),\cr
&c_2={1\over 16 \pi^2}({5\over 4}),\cr
&c_3={1\over 16 \pi^2}(-{1\over 2}),\cr
&c_4={1\over 16 \pi^2}(0).\cr}\eqno(3.25)$$
Now we have completely determined the terms in the two-fermion effective
Lagrangian with
up to one-derivative. This Lagrangian will be used in extracting
leading $m_t$ contributions to various physical processes.

Note the appearance of $1\over \epsilon '$ in some of the coefficients.
Their origins can be ascribed to subtraction of $\Omega _{ind}$, as well
as the necessity of conventional renormalizations.  When we take $L_{eff}=
L_{eff}^{tree}+ L_{eff}^{1 loop}+\cdot \cdot $ as a genuine Lagrangian and
perform a loop expansion, the $1\over \epsilon '$ due to subtraction of
$\Omega _{ind}$ to project out effective vertices will be cancelled out
automatically.  Loosely speaking, one may associate these $1 \over \epsilon '$
with a cutoff, the exact functional relation need not be specified,
which separates low momentum from high momentum $\approx
m_t$ physics.  This separation is, however, artificial, in the sense that it
arbitrary divides a momentum into high and low regions and is not even
the true scale of expansion, that being ${v^2 \over 2}$.  A physical result
cannot depend on such a demarcation.  Our construction procedure asserts that
the cancellation of these cutoff dependent terms
is automatically enforced within the low energy effective theory.  There
is no need to consult the high momentum sector for it to happen.  This is
how it works for any effective theory when the underlying theory is
renormalizable.  The organization for this to happen is through
Zimmermann oversubtraction identities and is intimately related to a known
result that
the renormalization of high dimensional operators depends on themselves and
the lower dimensional ones, but the renormalization of the low dimensional
operators requires no knowledge of the higher ones.  This point has
not so far been appreciated or fully understood by many users of effective
theories.  (The other $1\over \epsilon '$ are of course absorbed by
reparameterization.)

Before we close this section, we would like to make another comment on
our derivation of the
effective Lagrangian. As one has seen, we expand the generating
functional $W$ around the background fields chosen to be solutions of
nonlinear equation of motion rather than those which satisfy classical
equations
of motion. The difference between these two approaches will occur in the
treatment of external tree graphs. By expanding the generating functional
around a different set of classical fields,
the external tree graphs will be organized
in a different way, although the result after summing over all the diagrams
remains the same.  In appendix B, we shall
explicitly illustrate this point.

\noindent{\bf 4. APPLICATIONS}

In this section, we shall demonstrate the simplicity and usefulness of the
effective Lagrangian in locating the leading radiative corrections
to certain one-loop processes.
In the first two examples, we shall utilize the effective Lagrangian
to reproduce known results on the decays $Z\rightarrow b\bar{s}$ and
$Z\rightarrow b\bar{b}$.
The first process is a rare decay which does
not have any tree diagrams.  At the one-loop level,
the
amplitudes of both processes are sensitive to the top quark
mass. Some of the $m_t$-dependent terms are in fact proportional to
$m_t^2$, which can be obtained
quite easily by applying the effective Lagrangian derived in the last
section.  We shall demonstrate this momentarily.
In point of fact, if we want the complete subleading $O(m_t^0)$
effects, we should include terms with two and three derivatives in our
effective Lagrangian, acccording to Eq. (3.1).  We have not constructed
these three derivative terms, partly because the subleading terms are
unlikely to be detectable in flavor changing neutral current (FCNC)
processes in the near future.
As a third application, we shall use the effective Lagrangian to
evaluate $O(m_t^2)$ terms in the $\sigma b\bar{b}$ vertex.

\smallskip
\noindent{\bf (A) $Z\rightarrow b\bar{s}$}

It is a simple matter to generalize the effective Lagrangian in the
last section to include FCNC processes. By
replacing one of the $b$ quarks with a strange quark, we obtain an
effective Lagrangian which is suitable for the process in question.
{}From Eqs. (3.2) and (3.4) and shifting
the Higgs field, we obtain the following effective Lagrangian
for FCNC processes

$$\eqalign{L^{0 \partial}_{FCNC}&={g^2V^*_{ts}V_{tb}\over 32\pi^2
m_W^2}
[m_b(m_t^2+m_W^2)(-{1\over \epsilon^{\prime}}-1+ln({m_t^2\over
\kappa^2}))
\bar{s}_Lb_R]\cr
&+{g^2V^*_{ts}V_{tb}\over 32\pi^2 m_W^2}
[m_s(m_t^2+m_W^2)(-{1\over \epsilon^{\prime}}-1+ln({m_t^2\over
\kappa^2}))
\bar{s}_Rb_L]\cr
&+h.c.,\cr}\eqno(4.1)$$
$$\eqalign{L^{1\partial}_{FCNC}={g^2V^*_{ts}V_{tb}\over 16\pi^2
m_W^2}
[&m_t^2(-{1\over 4\epsilon{\prime}}-{3\over 8}+{1\over 4}
ln({m_t^2\over \kappa^2})\bar{s}_L{1\over i}\gamma\cdot D_nb_L\cr
&+m_bm_s(-{1\over 4\epsilon{\prime}}-{3\over 8}+{1\over 4}
ln({m_t^2\over \kappa^2})\bar{s}_R{1 \over i}\gamma\cdot D_nb_R\cr
&+2m_W^2(-{1\over 2\epsilon{\prime}}-{3\over 4}+{1\over 2}
ln({m_t^2\over \kappa^2})\bar{s}_L{1 \over i}\gamma\cdot D_nb_L\cr
&+{g\over 4 c_{\theta}}(m_t^2-2m_W^2(-{1\over 2\epsilon{\prime}}
+{1\over 2}ln({m_t^2\over
\kappa^2}))\bar{s}_L\gamma_{\mu}b_LZ^{\mu}\cr
&-{g\over 4 c_{\theta}}m_bm_s\bar{s}_R\gamma_{\mu}b_RZ^{\mu}]\cr
&+h.c.,\cr}\eqno(4.2)$$
where

$$D_n^{\mu}= \partial^{\mu}-ieQ_nA^{\mu}
-{ig\over c_{\theta}}(T_3^n-Q_ns_{\theta}^2)Z^{\mu},\eqno(4.3)$$
with $Q_n=-{1\over 3}$ and $T_3^n=-{1\over 2}$. $V_{ij}$ are the
relevant Cabbibo-Kobayashi-Maskawa matrix elements, and $c_{\theta}$ and
$s_{\theta}$ stand for the $cos\theta_w$ and $sin\theta_w$,
respectively.
Neither the higher-derivative operators nor the tree-level effective
Lagrangain $L_{eff}^{tree}$ generate any
additional $O(m_t^2)$-dependent terms.
In Eq. (4.2), the covariant derivative $D_n^{\mu}$ contains both the
neutral
$Z$ boson and the photon. Therefore, any term carrying $D_n^{\mu}$
will contribute to both $b\rightarrow s\gamma$ and $Z\rightarrow
b\bar{s}$.
These contributions, however, will be cancelled away
by the counter terms as we carry out the renormalization program
later on.
This then explains why the process
$b\rightarrow s\gamma$ does not contain any $m_t^2$
type of terms in its amplitude. However, for the decay
$Z\rightarrow b\bar{s}$, there are additional contributions from the
last
two terms in Eq. (4.2). These two terms
in fact arise from operators in Eq. (3.4) with coefficients $a_3$,
$a_4$,
$a_7$, $a_8$, $b_3$ and $b_4$. In particular,
the $O(m_t^2)$-dependent contributions for
$Z\rightarrow b\bar{s}$ exclusively come from the operator with the
coefficient $a_3$ after replacing $\phi \rightarrow {v \over \sqrt 2}$,
which clearly assigns the effects to SU(2) spontaneous symmetry breaking.
Let us now discuss the detail.

To perform the renormalization for FCNC processes, we need to first
compute
the flavor-changing self-energy diagrams. Let us keep
only those terms which are proportional to $m_t^2$.
{}From Eqs. (4.1) and (4.2), the
unrenormalized flavor-changing self-energy is given by
$$\Sigma_{sb}=C_1V_{ts}^*V_{tb}(m_b\bar{s}_Lb_R+m_s\bar{s}_Rb_L)
+C_2V_{ts}^*V_{tb}\bar{s}_L{1 \over i}\gamma\cdot\partial b_L,\eqno(4.4)$$
with
$$C_1={g^2m_t^2\over 32\pi^2m_W^2}(-{1\over \epsilon^{\prime}}-1
+ln({m_t^2\over \kappa^2}));\eqno(4.5a)$$
$$C_2={g^2m_t^2\over 32\pi^2m_W^2}(-{1\over 2\epsilon^{\prime}}-{3\over
4}
+{1\over 2}ln({m_t^2\over \kappa^2})).\eqno(4.5b)$$
\smallskip
The renormalized self-energy $\Sigma_{sb}^{ren}$ is obtained by
adding counter terms to $\Sigma_{sb}$ such
that, in momentum space,

$$\Sigma_{sb}^{ren}(P^2=-m_b^2)b=0;\eqno(4.6a)$$
$$\bar{s}\Sigma_{sb}^{ren}(P^2=-m_s^2)=0.\eqno(4.6b)$$
Such counter terms are generated by performing the following field
replacements in the original bare Lagrangian:

$$b_L\rightarrow Z_L^bb_L+V_{ts}V_{tb}^*\delta
Z_L^{bs}s_L,\eqno(4.7a)$$

$$b_R\rightarrow Z_R^bb_R+V_{ts}V_{tb}^*\delta
Z_R^{bs}s_R,\eqno(4.7b)$$

$$s_L\rightarrow Z_L^ss_L+V_{ts}^*V_{tb}\delta
Z_L^{sb}b_L,\eqno(4.7c)$$

$$s_R\rightarrow Z_R^ss_R+V_{ts}^*V_{tb}\delta
Z_R^{sb}b_R.\eqno(4.7d)$$
Upon adding these counter terms into $\Sigma_{sb}$, we obtain
$$\Sigma_{sb}^{ren}=V_{ts}^*V_{tb}(A_R\bar{s}_Lb_R+A_L\bar{s}_Rb_L
+B_L\bar{s}_L{1\over i}\gamma \cdot \partial b_L),\eqno(4.8)$$
with
$$A_R=(C_1-\delta Z_L^{bs*})m_b-\delta
Z_R^{sb}m_s,\eqno(4.9a)$$
$$A_L=(C_1-\delta Z_L^{sb})m_s-\delta
Z_R^{bs*}m_b,\eqno(4.9b)$$
and
$$B_L=C_2-\delta Z_L^{bs*}-\delta Z_L^{sb}.\eqno(4.9c)$$
\smallskip
{}From the renormalization
conditions in Eq. (4.6a) and (4.6b), one immediately concludes that
all of $A_R$, $A_L$ and $B_L$ are identically equal to zero. This
implies that the $O(m_t^2)$ renormalized self-energy $\Sigma_{sb}^{ren}$ also
vanishes.
That the renormalized flavor-changing self energy vanishes is not a
general
result, and is only true in the current case, because we have
now only included
contributions from dimension 4 operators in computing the self-energy
$\Sigma_{sb}$. Such contributions are completely  cancelled by
counter terms with the same dimension. Althougth $\Sigma_{sb}^{ren}$
could
in principle receive
contributions from higher dimension operators, these contributions,
however,
would not be proportional to $m_t^2$. Now that one has a vanishing
$\Sigma_{sb}^{ren}$, neither the process $Z\rightarrow b\bar{s}$
nor the one $b\rightarrow s\gamma$ would receive
any contribution from reducible diagrams. To find contributions from
the irreducible diagrams, let us first discuss the renormalization of
flavor-changing vertices $\Gamma_{sbZ}$ and $\Gamma_{sb\gamma}$. From
Eq. (4.2), one extracts
$$\eqalign{L_{FCNC}^{1\partial}\rightarrow
&C_2V_{ts}^*V_{tb}\bar{s}_L
\gamma_{\mu}b_L[-eQ_nA^{\mu}-{g\over
c_{\theta}}(T_3^n-Q_ns_{\theta}^2)
Z^{\mu}]\cr
&+C_3V_{ts}^*V_{tb}\bar{s}_L\gamma_{\mu}b_L({g\over
2c_{\theta}}Z^{\mu}),\cr}
\eqno(4.10)$$
where $C_2$ was defined in Eq. (4.5b) and $C_3$ is given by
$$C_3={g^2m_t^2\over 32\pi^2m_W^2}.\eqno(4.11)$$
The second term in Eq. (4.10) in fact
came from the operator with coefficient
$a_3$ in the last section. Now it is clear that Eq. (4.10) will give
renormalization
to the bare vertices $\Gamma_{sbZ}$ and $\Gamma_{sb\gamma}$. The
counter
terms for renormalizing these vertices are again generated by field
replacements laid out in Eqs. (4.7a)-(4.7d). Upon adding the counter
terms,
the resulting renormalized vertex $\Gamma_{\bar{s}bZ(\gamma)}^{ren}$
reads
$$\eqalign{\Gamma_{\bar{s}bZ(\gamma)}^{ren}&=B_LV_{ts}^*V_{tb}\bar{s}
_L
\gamma_{\mu}[-eQ_nA^{\mu}-{g\over
c_{\theta}}(T_3^n-Q_ns_{\theta}^2)
Z^{\mu}]b_L\cr
&+C_3V_{ts}^*V_{tb}\bar{s}_L\gamma_{\mu}b_L({g\over
2c_{\theta}}Z^{\mu}).\cr}
\eqno(4.12)$$

Now the first term in the above equation vanishes, due to $B_L=0$
according to
Eq. (4.9c). The immediate consequence for this is
that the $\bar{s}bA_{\mu}$
vertex would not contain any $O(m_t^2)$ terms.
As a result, the decay $b\rightarrow
s\gamma$ is shown to be free of $m_t^2$ contributions.
However, for $Z\rightarrow
b\bar{s}$, the additional contribution coming from
the second term in Eq. (4.12) is clearly
$O(m_t^2)$. With $C_3$ given by Eq. (4.11), we arrive at

$$\Gamma_{\bar{s}bZ}^{ren}={g^3\over 16\pi^2 m_W^2c_{\theta}}
V_{ts}^*V_{tb}({1\over
4}m_t^2)\bar{s}_L\gamma_{\mu}b_LZ^{\mu},\eqno(4.13)$$
which is the total amplitude for the decay $Z\rightarrow b\bar{s}$
as reducible diagrams were shown earlier to have no contributions. Our result
presented
in Eq. (4.13) reproduces the one obtained by Inami and Lim [10].

\bigskip
\noindent{\bf (B) $Z\rightarrow b\bar{b}$}
\smallskip
In computing radiative corrections to the decay $Z\rightarrow
b\bar{b}$,
One needs again to specify a renormalization scheme. We shall employ the
on-shell
scheme with input parameters chosen to be the electric charge $e$,
the
physical mass of vector bosons $m_Z$ and $m_W$, the physical mass of
fermions
$m_f$, and the physical mass of Higgs $m_H$. The charged boson mass
$m_W$
will later be related to the Fermi constant $G_F$. To
generate
counter terms, we first write the tree-level
$Zb\bar{b}$ vertex in terms of bare
quantities:
$$\eqalign{L_{tree}&=({\sqrt 2}G_F^0)^{1\over 2}
m_Z^0\bar{b}_0\gamma^{\mu}
[-2Q_n(s_{\theta}^0)^2+T_3^n(1-i\gamma_5)]b_0Z_{\mu}^0\cr
&=({\sqrt 2})^{1\over 2}Z_{\mu}^0\bar{b}_0\gamma^{\mu}
[v_b^0-ia_b^0\gamma_5]b_0.\cr}\eqno(4.14)$$
where
$$v_b^0=(G_F^0)^{1\over
2}m_Z^0(T_3^n-2Q_n(s_{\theta}^0)^2),\eqno(4.15)$$
$$a_b^0=(G_F^0)^{1\over 2}m_Z^0T_3^n.\eqno(4.16)$$
Now each bare quantity will be expressed in terms of the renormalized
one plus the counter term. For $v_b^0$ and $a_b^0$, we have
$$v_b^0=v_b+\delta v_b; \ \ \ a_b^0=a_b+\delta a_b,\eqno(4.16)$$
The counter terms $\delta v_b$ and $\delta a_b$ are given by

$$\delta v_b={1\over 2}v_b({\delta G_F\over G_F}+{\delta
M_Z^2\over
M_Z^2})-2Q_ns_{\theta}^2(G_F)^{1\over 2}m_Z{\delta
s_{\theta}^2\over
s_{\theta}^2},\eqno(4.17)$$
$$\delta a_b={1\over 2}a_b({\delta G_F\over G_F}+{\delta
M_Z^2\over
M_Z^2}),\eqno(4.18)$$
The $G_F$ in Eqs. (4.17) and (4.18) are chosen to be the Fermi
constant. This choice then fixes the counter term $\delta G_F$.
Since
$$G_F^0={\pi \alpha^0\over {\sqrt 2}(s_{\theta}^0)^2
(m_W^0)^2},\eqno(4.19)$$
We have
$$\eqalign{G_F^0&=G_F(1+{\delta G_F\over G_F})\cr
&=G_F(1-\Delta r +{\delta \alpha\over \alpha}-{\delta m_W^2\over
m_W^2}
-{\delta s_{\theta}^2\over s_{\theta}^2}),\cr}\eqno(4.20)$$
The correction term $\Delta r$ appears because of the following
relation
among physical parameters
$$G_F={\pi \alpha\over {\sqrt 2}s_{\theta}^2m_W^2(1-\Delta r)},
\eqno(4.21)$$
We do not write down the complete expression for
$\Delta r$, but only to isolate its leading terms. Thus, we write
$$\Delta r=-{c_{\theta}^2\over s_{\theta}^2}\Delta \rho+\Delta
r_{rem},
\eqno(4.22)$$
Here $\Delta \rho$ is the one loop correction to the $\rho$
parameter.
$\Delta r_{rem}$ is the remaining contributions which contain no
$O(m_t^2)$ terms. From now on, all the non-$O(m_t^2)$ terms will be
neglected, which means that we can discard
the term ${\delta \alpha\over \alpha}$ in Eq. (4.20).
To further simplify the equation, one can make use of the relation:
$${\delta s_{\theta}^2\over s_{\theta}^2}={c_{\theta}^2\over
s_{\theta}^2}
({\delta m_Z^2\over m_Z^2}-{\delta m_W^2\over m_W^2})=
{c_{\theta}^2\over s_{\theta}^2}\Delta \rho.\eqno(4.23)$$

{}From Eqs. (4.17), (4.18), (4.20), (4.22) and (4.23), we arrive at
$$\delta v_b={1\over 2}v_b\Delta \rho-2Q_nG_F^{1\over 2}
m_Zc_{\theta}^2\Delta \rho,\eqno(4.24)$$
$$\delta a_b={1\over 2}a_b\Delta \rho.\eqno(4.25)$$
\smallskip
Besides contributions from the counter terms, there are also other
contributions from one-loop vertex diagrams as well as wave function
renormalizations of the external lines.
For the process $Z\rightarrow b\bar{b}$, it is
easy to see by the power counting argument that only the $b$
quark wave function renormalization could possibly
give rise to $O(m_t^2)$ terms. To compute its effect, let us review
some basic steps in
the renormalization of chiral fermions. Taking the $b$
quark as an example, we may parametrize its
self-energy as follows:
$$\Sigma_b( k\!\!\!/\ )= k\!\!\!/\
(A(k^2)-iB(k^2)\gamma_5)+m_b^0C(k^2),
\eqno(4.26)$$
where $m_b^0$ is the bare mass of the $b$ quark. Now the dressed
$b$ quark propagator becomes
$$S_b(k\!\!\!/\ )={1\over (2\pi)^4 i}{1\over  k\!\!\!/\  +m_b^0
-\Sigma( k\!\!\!/\ ) }.\eqno(4.27)$$
Expanding $S_b$ around $k^2=-m_b^2$, we obtain
$$S_b\approx {1\over (2\pi)^4 i}{(- k\!\!\!/\ +m_b)Z-i k\!\!\!/\
\gamma_5
\underline B\over k^2+m_b^2},\eqno(4.28)$$
where $\underline B$ is a shorthand notation for $B(k^2=-m_b^2)$,
$m_b$ is the physical mass given by
$$m_b^2=(m_b^0)^2(1+2{\underline A}-2{\underline C}),\eqno(4.29)$$
and
$$Z=1+{\underline
A}+2m_b^2(\underline{C'}-\underline{A'}).\eqno(4.30)$$
As before, all the underlined quantities are evaluated at
$k^2=-m_b^2$.
Furthermore, $A'$ and $C'$ are defined as

$$A'={\partial A\over \partial k^2}; \ \ C'={\partial C\over \partial
k^2}.
\eqno(4.31)$$
\smallskip
In the decay $Z\rightarrow b\bar{b}$,
the wave function for an outgoing $\bar{b}$ quark
is $v(-k,s)$, which may be
decomposed into left-handed and right-handed components:
$$v_L={1-i\gamma_5\over 2}v(-k,s); \ \ v_R={1+i\gamma_5\over 2}v(-k,s).
\eqno(4.32)$$
By the reduction formula, the renormalized
wave functions $v^{ren}(-k,s)$ should be taken as
$$\eqalign{v^{ren}(-k,s)&={1\over 2m_b\sqrt{Z_L}}[- k\!\!\!/\
(Z_LP_L
+ Z_RP_R)+m_b Z]v_L(-k,s)\cr
&+{1\over 2m_b\sqrt{Z_R}}[- k\!\!\!/\ (Z_LP_L
+ Z_RP_R)+m_b Z]v_R(-k,s)\cr
&={1\over 2}(Z+1-i\underline{B}\gamma_5)v(-k,s),\cr}\eqno(4.33)$$
where $Z_L=Z-\underline B$, and $Z_R=Z+\underline B$. In deriving the
above
equation, we have set $\sqrt{Z_LZ_R}=Z$ since we are only concerned
with contributions up to one-loop order.   Now the result stated in
Eq. (4.33) can also be acquired by shifting the $b$ quark field
according to
$$b\rightarrow {1\over 2}(Z+1-i{\underline B}\gamma_5)b,\eqno(4.34)$$
and hence
$$\bar{b}\rightarrow \bar{b}{1\over 2}(Z+1+i{\underline B}\gamma_5).
\eqno(4.35)$$
Such field-shiftings generate the following effective couplings:
$$\eqalign{L_{wr}&=({\sqrt 2}G_F)^{1\over 2}m_Z\bar{b}
[{1\over 2}(Z+1+i{\underline B}\gamma_5)-1]\gamma_{\mu}
(-2Q_ns_{\theta}^2+T_3^n(1-i\gamma_5))bZ^{\mu}\cr
&+({\sqrt 2}G_F)^{1\over 2}m_Z\bar{b}\gamma_{\mu}
(-2Q_ns_{\theta}^2+T_3^n(1-i\gamma_5))[{1\over 2}(Z+1-i{\underline
B}\gamma_5)-1]
bZ^{\mu}.\cr}\eqno(4.36)$$
To compute the unknown parameters in the equation above, we need to
extract the $b$ quark self-energy from the effective Lagrangian.
{}From Eqs. (3.2) and (3.4), we can easily read off

$$\eqalign{\Sigma_b( k\!\!\!/\ )={g^2m_t^2\over 32\pi^2 m_W^2}
&[(-{1\over 4\epsilon^{\prime}}-{3\over 8}+{1\over 4}ln({m_t^2\over
\kappa^2}))
 k\!\!\!/\ (1-i\gamma_5)\cr
&+m_b(-{1\over \epsilon^{\prime}}-1+ln({m_t^2\over
\kappa^2})].\cr}\eqno(4.37)$$
This yields
$$A=B={g^2m_t^2\over 32\pi^2 m_W^2}
(-{1\over 4\epsilon^{\prime}}-{3\over 8}+{1\over 4}ln({m_t^2\over
\kappa^2})),
\eqno $$
and
$$C={g^2m_t^2\over 32\pi^2 m_W^2}
(-{1\over \epsilon^{\prime}}-1+ln({m_t^2\over \kappa^2})).
\eqno(4.38)$$
Eqs. (4.36) and (4.38) together give
$$\eqalign{L_{wr}=&{g^2m_t^2\over 32\pi^2m_W^2}
(-{1\over 4\epsilon^{\prime}}-{3\over 8}+{1\over 4}ln({m_t^2\over
\kappa^2}))
\times\cr
&({\sqrt 2}G_F)^{1\over 2}m_Z\bar{b}\gamma_{\mu}
(-2Q_ns_{\theta}^2+2T_3^n)(1-i\gamma_5)bZ^{\mu}.\cr}\eqno(4.39)$$
Finally, from Eqs. (3.2) and (3.4), we obtain the following vertex
corrections
$$\eqalign{L_{ver}=&-{g^2m_t^2\over 32\pi^2m_W^2}
(-{1\over 4\epsilon^{\prime}}-{3\over 8}+{1\over 4}ln({m_t^2\over
\kappa^2}))
\times\cr
&({\sqrt 2}G_F)^{1\over 2}m_Z\bar{b}\gamma_{\mu}
(-2Q_ns_{\theta}^2+2T_3^n)(1-i\gamma_5)bZ^{\mu}\cr
&+{g^3m_t^2\over
64\pi^2c_{\theta}m_W^2}\bar{b}_L\gamma_{\mu}b_LZ^{\mu}.\cr}
\eqno(4.40)$$
The first term in the vertex corrections cancells the contribution
$L_{wr}$
from the $b$ quark wave function renormalization. The second term,
which
we now denote as $L_{ver}^{\prime}$, can be written as

$$L_{ver}^{\prime}=\Delta \rho_{ver}
({\sqrt 2}G_F)^{1\over 2}m_Z\bar{b}\gamma_{\mu}
T_3^n(1-i\gamma_5)bZ^{\mu},\eqno(4.41)$$
with

$$\Delta \rho_{ver}=-{g^2m_t^2\over 32\pi^2m_W^2}.\eqno(4.42)$$
The contribution $L_{ver}^{\prime}$ in fact has the same origin as
the
second term in Eq. (4.10). They both come from the operator with
coefficient
$a_3$ in the last section.
Now we are ready to obtain the leading radiative corrections to the
process $Z\rightarrow b\bar{b}$.
For convenience, let us denote the
original $\Delta \rho$ by $\Delta \rho_{se}$

$$\Delta \rho_{se}={3g^2m_t^2\over 64\pi^2m_W^2}.\eqno(4.43)$$
Adding Eqs. (4.24), (4.25), (4.41), (4.42) and (4.43)
to the tree-level coupling in Eq. (4.14), we obtain
$$\eqalign{L_{Zb\bar{b}}=&(1+{1\over 2}\Delta \rho_{se}+\Delta
\rho_{ver})
({\sqrt 2}G_F)^{1\over 2}m_Z\bar{b}\gamma_{\mu}
T_3^n(1-i\gamma_5)bZ^{\mu}\cr
&+(1+{1\over 2}\Delta \rho_{se})
({\sqrt 2}G_F)^{1\over 2}m_Z\bar{b}\gamma_{\mu}
[-2Q_ns_{\theta}^2(1+{c_{\theta}^2\over s_{\theta}^2}\Delta
\rho_{se})]
bZ^{\mu}.\cr }\eqno(4.44)$$
Our result here again reproduces the one
obtained previously by direct calculations [11].
\bigskip
\noindent{\bf (C) $\sigma \rightarrow b\bar{b}$}
\smallskip
To further demonstrate the computational efficiency provided by
the effective Lagrangian, we now
evaluate the vertex of $\sigma \rightarrow b\bar{b}$. In view of the
likelihood that $\sigma$ (Higgs) may be rather heavy, the result
should be
used when $\sigma$ is virtual.
\smallskip
Throughout this discussion, we take $m_t$ as the largest scale in the
problem
and concentrate only on the leading corrections
proportional to $G_Fm_t^2$. The effective vertex for
$H\rightarrow b \bar{b}$ to the one-loop order can be written as
$$L^{\sigma b\bar{b}}=-2^{{1\over 4}}
G_F^{{1\over 2}}m_b(1+\Delta T)
\sigma
\bar{b}b,\eqno(4.45)$$
where we use $\sigma$ to represent the Higgs field, and $\Delta T$ to
denote
one-loop contributions arising from vertex diagrams and
counter terms. For convenience, we write
$$\Delta T=\Delta T_{ver.}+ \Delta T_{ct.}.\eqno(4.46)$$
By neglecting non-$O(m_t^2)$ contributions, it is easy to show that
$$\Delta T_{ct.}=({\delta m_b\over m_b}-{1\over 2}{\delta m_W^2\over
m_W^2}
+{1\over 2}\delta Z^{\sigma}+\delta Z^b),\eqno(4.47)$$
where $\delta Z^{\sigma}$ and $\delta Z^b$ denote wave function
renormalizations needed for external Higgs and $b$ quarks,
respectively.
It is useful to separate $\Delta T$ into {\it universal} and
{\it non-universal}
corrections, where the former one is defined as the correction which
occurs
in any two-fermion decay mode of the Higgs boson. For instance, the
effective
vertex for $H\rightarrow \tau^+\tau^-$ to the one-loop order is given
by
$$L^{\sigma \tau^+\tau^-}=-2^{{1\over 4}}G_F^{{1\over
2}}m_{\tau}(1+\Delta
T_{universal})\sigma \bar{\tau}\tau,\eqno(4.48)$$
where
$$\Delta T_{universal}= -{1\over 2}{\delta m_W^2\over m_W^2}+{1\over
2}\delta
Z^{\sigma}.\eqno(4.49)$$
Since our result is useful only when $\sigma$ is virtual, it
appears as an internal propagator. In this sense, we are actually
computing the off-shell correction coming from the dressed
Higgs-boson
propagator.  Now the dressed Higgs-boson
propagator
reads:
$$\Delta_{\sigma}(P^2)={1\over (2\pi)^4i}{1\over P^2+m_{\sigma}^0
-\Sigma_{\sigma}(P^2)},\eqno(4.50)$$
where $\Sigma_{\sigma}(P^2)$ is the Higgs-boson self-energy. To
compute
the off-shell corrections, we perform the following expansion:
$$\Sigma_{\sigma}(P^2)=\Sigma_{\sigma}(P^2=-m_{\sigma}^2)
+{\partial \Sigma_{\sigma}\over \partial P^2}|_{P^2=-m_{\sigma}^2}
(P^2+m_{\sigma}^2)+ \ \cdots .\eqno(4.51)$$
Here we have neglected higher order terms in the expansion since they
contain no $O(m_t^2)$ contributions. Also $m_{\sigma}$ is the
physical
Higgs-boson mass to be defined shortly.
Substituting Eq. (4.51) into Eq. (4.50), the dressed Higgs-boson
propagator
becomes
$$\Delta_{\sigma}(P^2)={{1\over (2\pi)^4i}{(1+{\partial \Sigma \over
\partial P^2}|_{P^2=-m_{\sigma}^2})\over
P^2+m_{\sigma}^2}},\eqno(4.52)$$
with
$$m_{\sigma}^2=(m_{\sigma}^0)^2-\Sigma_{\sigma}(P^2=-m_{\sigma}^2).
\eqno(4.53)$$
\smallskip
{}From Eq. (4.52), it is clear that the off-shell correction in
our current approximation precisely corresponds to the usual
wave function renormalization for the Higgs field.
To obtain this correction, we resort to
Eq. (I-7.37) and have
$${1\over 2}\delta Z^\sigma={1\over 2}{\partial \Sigma_{\sigma}\over
\partial P^2}
|_{P^2=-m_{\sigma}^2}={G_Fm_t^2\over
4
{\sqrt 2}\pi^2}(-{3\over 2\epsilon^{\prime}}
+{21\over 12}+{3\over 2}ln({m_t^2\over \kappa^2})),\eqno(4.54)$$
Note that we have re-expressed ${H^2\over 16\pi^2}$ as ${G_Fm_t^2\over
4
{\sqrt 2}\pi^2}$. Also, please note that a color factor $N_c=3$
should be
added to results given in I.
The mass renormalization of gauge bosons ${\delta m_W^2
\over m_W^2}$ can be easily computed by refering
to the following operator, again from Eq.(I-7.37)
$$a (D_{\mu}\phi)^{\dagger}(D^{\mu}\phi),\eqno(4.55)$$
with
$$a={N_c\over 16\pi^2}(H^2+h^2)(-{1\over
\epsilon^{\prime}}+ln({H^2\bar{\phi}
\phi \over \kappa^2})).\eqno(4.56)$$
Selecting only $O(m_t^2)$ contributions, we find
$${\delta m_W^2\over m_W^2}={G_Fm_t^2\over 2{\sqrt 2}\pi^2}
(-{3\over 2\epsilon^{\prime}}+{3\over 2}ln({m_t^2\over \kappa^2})).
\eqno(4.57)$$
{}From Eqs. (4.54) and (4.57), we obtain
$$\Delta T_{universal}={7G_F^2m_t^2\over
16\sqrt{2}\pi^2}.\eqno(4.58)$$
This agrees with the results by Dabelstein and Hollik and by Kniehl [16].
We next look into the {\it non-universal correction}. Such a
correction exists
in the decay mode $H\rightarrow b\bar{b}$. Following Eqs. (4.46) and
(4.47),
we write
$$\Delta T_{non-universal}=\Delta T_{ver.} +{\delta m_b\over m_b}
+\delta Z^b.\eqno(4.59)$$
For $\Delta T_{ver.}$, we obtain from the effective potential in
Eq.(3.2)
and the one derivative operator with the coefficient $a_2$ in Eq.
(3.4)
$$\Delta T_{ver.}=-{G_Fm_t^2\over 4{\sqrt 2}\pi^2}(-{1\over
\epsilon^{\prime}}
+{1\over 2}+ln({m_t^2\over \kappa^2})).\eqno(4.60)$$
We would like to point out that
the Higgs field could come from the nonlocal operator
$ln({\bar{\phi}\phi\over \kappa^2})$ due to the following expansion:
$$ln({\bar{\phi}\phi\over \kappa^2})=ln({(\sigma +v)^2+\vec
\pi^2\over
2\kappa^2})
={2\sigma \over v}+ \ \cdots.\eqno(4.61)$$
Finally, we turn to the wave function and the mass renormalization of
the
b-quark.  We use Eqs.(4.29) and (4.30) to obtain
$${\delta m_b\over m_b}+\delta
Z^b=C(k^2=-m_b^2)+2m_b^2(C'(k^2=-m_b^2)-
A'(k^2=-m_b^2)).\eqno(4.62)$$
By the power counting rule stated in our Eq. (3.1), only the first
term
$C(k^2=-m_b^2)$ contains the $O(G_Fm_t^2)$ contribution. This
contribution
can be found in our Eq. (4.38):
$$C(k^2=-m_b^2)={G_Fm_t^2\over 4{\sqrt 2}\pi^2}(-{1\over
\epsilon^{\prime}}
-1+ln({m_t^2\over \kappa^2})).\eqno(4.63)$$
By Eqs. (4.58), (4.59), (4.60), (4.62) and (4.63), we obtain the
total
$O(G_Fm_t^2)$ correction to $H\rightarrow b\bar{b}$
$$\Delta T={G_Fm_t^2\over 16{\sqrt 2}\pi^2},\eqno(4.64)$$
which agrees with the results by previous authors [16] in direct
calculation.

{\bf 5. Conclusion }
\smallskip
We have explicitly constructed one loop effective Lagrangian to
account for virtual top quark effects, in processes where there are
two external light quarks and an arbitrary number of light bosons.  We
have reproduced some well-known results to demonstrate the universal
nature of our method.  We have found it very simple and systematic to
apply for this purpose.
\smallskip
On a more formal level, we have shown that no new infinities are
introduced in our calculations, although higher dimensional operators
are used.  Together with our discussion before, we want to reiterate
that the effective theory so constructed by us are fully equivalent
to the full theory for low energy physics.  They each result in
an analytic renormalizable loop expansion.  The neglected terms are
indeed under control and negligible.

\bigskip
\noindent
{\bf Acknowledgements }

\smallskip
We would like to thank C. P. Yuan for his interest and discussions. We also
thank A. Dabelstein and B. A. Kniehl for discussions on electroweak corrections
to $H\rightarrow b\bar{b}$.
Both G.L.L. and H.S. would like to thank the Particle Theory Group of
the University of Michigan for their hospitality, where a part of this work
was done.  G.L.L. is supported in part by the National Science Council
of the Republic of China, under the contract number
NSC82-0208-M-001-011Y.  Y.P.Y. has been partially supported by the U.
S. Department of Energy throughout the investigation of this work.
\endpage
\noindent{\bf  APPENDIX A: TREE-LEVEL $b-\phi$ SCATTERING AMPLITUDE \break
AND $L_{eff}^{tree}$.}

As discussed in section 3, there exist two methods to subtract the quantity
$\Omega_{ind.}^{1loop}$ from $\Omega_{1LPI}^{1loop}$ in order to
construct the 1-loop effective Lagrangian $L_{eff}^{1loop}$.
The first method is to compute $\Omega_{ind.}^{1loop}$ directly from the
tree-level effective Lagrangian $L_{eff}^{tree}$ and have the result
subtracted from $\Omega_{1LPI}^{1loop}$. The alternative method is
to identify contributions of $\Omega_{ind.}^{1loop}$ from each
individual 1-loop diagram to be evaluated for determining
$\Omega_{1LPI}^{1loop}$. For this purpose, we have defined a $\tau$
-operation which acts on each top-quark propagator inside the 1-loop
diagram under consideration. Such operation will shrink top-quark
propagators and create effective vertices which are precisely identical
to those given by $L_{eff}^{tree}$. Through this, we have an equivalent
way to obtain $\Omega_{ind.}^{1loop}$. We then subtract the contributions
induced by the $\tau$-operation from the original diagrams and the result
will give the 1-loop effective Lagrangian $L_{eff}^{1loop}$.
Now the prescription for the $\tau$-operation should be fixed in such a
way that it will create an identical set of effective vertices as
the one given by $L_{eff}^{tree}$. To find the correct prescription,
let us consider the tree-level $b-\phi$ scattering amplitude. There
are four diagrams contributing to this process as shown in Fig. 2.
To match with the result arising from $L_{eff}^{tree}$, we shall neglect
contributions down by ${1\over H}$ or higher.
The amplitude for the first diagram reads:
$$\eqalign{I_a&=-h^2\int d^4xd^4y \bar{b}_R(x)\phi^-(x)<T(T_L(x)
\bar{T}_L(y))>b_R(y)\phi^+(y)\cr
&=-h^2\int d^4xd^4y \int {d^4p\over (2\pi)^4}e^{ip(x-y)}
\bar{b}_R(x)\phi^-(x)<T(T_L\bar{T}_L(y+i{\overrightarrow{\partial}\over
\partial p}))>\cr
& \ \ \ b_R(y)\phi^+(y).\cr}\eqno(A-1)$$
Now the operation $\tau$ will act on the top-quark propagator
$<T(T_L(x)\bar{T}_L(y))>$ and extract its leading terms which, when combined
with other terms in Eq. (A.1), should give rise to an amplitude which is
at least $O(H^0)$. Since
neither one of the coupling constants in $I_a$ is proportional to
$H$, even the most dominant term in $<T(T_L(x)\bar{T}_L(y))>$
gives no $O(H^0)$ result for the amplitude. We hence take
$$\tau <T(T_L(x)\bar{T}_L(y))>=0.\eqno(A-2)$$
The second diagram has the amplitude
$$\eqalign{I_b&=Hh\int d^4xd^4y \bar{b}_L(x)\phi^-(x)<T(T_R(x)
\bar{T}_L(y))>b_R(y)\phi^+(y)\cr
&=Hh\int d^4xd^4y \int {d^4p\over (2\pi)^4}e^{ip(x-y)}
\bar{b}_L(x)\phi^-(x)<T(T_R\bar{T}_L(y+i{\overrightarrow{\partial}\over
\partial p}))>\cr
& \ \ \ b_R(y)\phi^+(y).\cr}\eqno(A-3)$$
To obtain the leading terms, we have the operation $\tau$ acting on
the top quark propagator according to
$$\tau <T(T_R\bar{T}_L(y+i{\overrightarrow{\partial}\over
\partial p}))>={-iH\phi^{0\dagger}(y)\over H^2u(y)}R,\eqno(A-4)$$
with $u=\phi^{0\dagger}\phi^0$, and $R$ being the right-handed
projection operator.
In deriving the above equation, we
need to obtain first an expression for the top-quark propagator in the
form of derivative expansions and then have it acted by $\tau$.
We shall not give the details here but direct the reader to
an example shown in Eq. (I-5.12).
With the prescription given by Eq. (A-4), the leading terms in $I_b$
are found to be
$$I_b\approx -ih\int d^4x {\bar{b}_L(x)b_R(x)\phi^+(x)\phi^-(x)\over
\phi^0(x)}.\eqno(A-5)$$
This precisely corresponds to one of the effective vertices generated
by $L_{eff}^{tree}$ in Eq. (2.12).
Now the coordinate representation of Eq. (A-4) can be derived by using
$$<T(T_R(x)\bar{T}_L(y))>=\int {d^4p\over (2\pi)^4}e^{ip(x-y)}
<T(T_R\bar{T}_L(y+i{\overrightarrow{\partial}\over
\partial p}))>.\eqno(A-6)$$
Therefore,
$$\tau<T(T_R(x)\bar{T}_L(y))>={-i\delta^4(x-y)\over H\phi^0(x)}R.\eqno(A-7)$$
Similarly, one obtains
$$\tau<T(T_L(x)\bar{T}_R(y))>={-i\delta^4(x-y)\over H\phi^{0\dagger}(x)}L,
\eqno(A-8)$$
from the third diagram.
Finally, let us consider the fourth diagram in Fig. 2.
The amplitude for this diagram reads:
$$\eqalign{I_d&=-H^2\int d^4xd^4y \bar{b}_L(x)\phi^-(x)<T(T_R(x)
\bar{T}_R(y))>b_L(y)\phi^+(y)\cr
&=-H^2\int d^4xd^4y \int {d^4p\over (2\pi)^4}e^{ip(x-y)}
\bar{b}_L(x)\phi^-(x)<T(T_R\bar{T}_R(y+i{\overrightarrow{\partial}\over
\partial p}))>\cr
& \ \ \ b_L(y)\phi^+(y).\cr}\eqno(A-9)$$
{}From Eq. (I-5.12), we take
$$\eqalign{\tau <T(T_R\bar{T}_R(y+i{\overrightarrow{\partial}\over
\partial p}))>&={i\gamma \cdot p\over H^2u(y)}L+{\gamma \cdot \partial u(y)
\over H^2u^2(y)}L\cr
&-{(\gamma \cdot \partial \phi^{0\dagger}(y))\phi^0(y)\over H^2u^2(y)}L.\cr}
\eqno(A-10)$$
This prescription gives
$$I_d\approx -\int d^4x {\bar{b}_L(x)\phi^-(x)\over \phi^0(x)}
\gamma \cdot \partial ({b_L(x)\phi^+(x)\over \phi^{0\dagger}(x)}),
\eqno(A-11)$$
which again coincides with one of the effective vertices generated by
$L_{eff}^{tree}$ in Eq. (2.12). In the coordinate space, the
prescription in Eq. (A-10) becomes
$$\eqalign{\tau<T(T_R(x)\bar{T}_R(y))>=&{\gamma_{\mu}\phi^{0\dagger}(y)
(\partial_{\mu}^y \phi^0(y))\over H^2u^2(y)}
\delta^4(x-y)L\cr
&+{\gamma_{\mu}\over H^2u(y)}\partial_{\mu}^x
\delta^4(x-y)L\cr}.\eqno(A-12)$$

\noindent{\bf  APPENDIX B: EQUIVALENCE OF DIFFERENT BACKGROUND-FIELD
EXPANSIONS.}

As pointed out at the end of section 3, we have chosen the classical
fields in the background-field expansions to be solutions of nonlinear
equations of motion. This is different from the conventional approach where
solutions of classical equations of motion are chosen to be the background
fields. Expanding the generating functional around the nonlinear fields
has the advantage of better organizing in powers of $m_t$ of
the effective action in loop expansion. This technique has been exposed
before [3] and we do not intend to repeat the discussions here.
The main purpose for this section is to illustrate that different choices of
background fields correspond to different treatments for external lines.
As shown in Fig. 7, diagrams (7a) and (7b) are derived from the approach
where the generating functional $W$ are expanded around fields which
are solutions of
classical equations of motion. Diagrams (7c) and (7d), on the other
hand, are as a whole corresponding diagrams given by expanding $W$ around
nonlinear
fields. The helicity configurations of top-quark propagators are indicated
explicitly in the diagrams, and the blob in each diagram signify arbitrary
configurations with the restriction that any of such configuration must be
one-particle-irreducible. In other words, we shall illustrate the matching
of a special class
of $1LPI$ diagrams which contain only one tree-level top-quark propagator by
showing that contributions from diagrams (7a) and (7b) are identical to
those from diagrams (7c) and (7d). First of all, let us consider
diagram (7a) with a top-quark propgator $<T(T_R(x)\bar{T}_R(y))>$. Its
amplitude can be written as
$$I_{a1}=\int d^4x d^4yF(x)<T(T_R(x)\bar{T}_R(y))>
(-H)\phi^+b_L(y),\eqno(B-1)$$
where $F(x)$ denotes contributions from the blob.
The amplitude $I_{a1}$ will be matched with part of the contributions from
diagram (7d). This part of contributions can be extracted
by taking the top-quark propagator
in diagram (7d) to be $<T(T_R(x)\bar{T}_L(y))>$ and discarding the piece
${h\phi^+b_R\over \phi^0}$ from the structure $t_R^{(1)}$ appearing in the
external vertex. To be
precise, let us name this amplitude as $I_{d1}$ and we have
$$I_{d1}=\int d^4x d^4yF(x)<T(T_R(x)\bar{T}_L(y))>(i\gamma \cdot
\partial_y t_L^{(0)}(y)).\eqno(B-2)$$
Performing $integration \ by \ parts$ and using
$$<T(T_R(x)\bar{T}_L(y))>\gamma \cdot \overleftarrow{\partial_y}
=<T(T_R(x)\bar{T}_L(y)\gamma \cdot \overleftarrow{\partial_y})>,\eqno(B-3)$$
we obtain
$$I_{d1}=\int d^4x d^4y  F(x)
<T(T_R(x)\bar{T}_L(y){1\over i}\gamma \cdot \overleftarrow{\partial_y})>
t_L^{(0)}(y) .\eqno(B-4)$$
Now applying the equation of motion
$$\bar{T}_L(y){1 \over i}\gamma \cdot \overleftarrow{\partial_y}
=H\phi^{0\dagger}(y)\bar{T}_R(y),\eqno(B-5)$$
and the constraint $t_L^{(0)}=-{b_L\phi^+\over \phi^{0\dagger}}$, one can
easily show that $I_{d1}=I_{a1}$.
For the second set of diagrams,
we take the top-quark propagator
in diagram (7a) to be $<T(T_L(x)\bar{T}_R(y))>$,
which gives rise to an amplitude
$$I_{a2}=\int d^4x d^4yF(x)<T(T_L(x)\bar{T}_R(y))>
(-H)\phi^+b_L(y),\eqno(B-6)$$
The same amplitude can be generated from diagram (7d)
with the top-quark propagator given by $<T(T_L(x)\bar{T}_L(y))>$
and the vertex factor $t_R^{(1)}$ taken as
${i\gamma \cdot \partial t_L^{(0)}\over \phi^0}$.
To show that this is indeed true, let us follow the
above prescription to obtain an amplitude
$$I_{d2}=\int d^4x d^4yF(x)<T(T_L(x)\bar{T}_L(y))>(i\gamma \cdot
\partial_y t_L^{(0)}(y)).\eqno(B-7)$$
Performing $integration \ by \ parts$ and using
$$\eqalign{&<T(T_L(x)\bar{T}_L(y))>{1\over i}\gamma
\cdot \overleftarrow{\partial_y}\cr
&=<T(T_L(x)\bar{T}_L(y){1\over i}\gamma
\cdot \overleftarrow{\partial_y})>+i\delta^4(x-y)L,\cr}
\eqno(B-8)$$
we arrive at
$$\eqalign{I_{d2}&=\int d^4x d^4yF(x)<T(T_L(x)\bar{T}_L(y){1\over i}\gamma
\cdot \overleftarrow{\partial_y})>t_L^{(0)}\cr
& \ \ +i\int d^4x d^4y F(x) \delta^4(x-y)t_L^{(0)}(y).\cr}\eqno(B-9)$$
By Eq. (B-5), we obtain
$$\eqalign{I_{d2}&=\int d^4x d^4yF(x)<T(T_L(x)\bar{T}_R(y))>(-H)
\phi^+(y)b_L(y)\cr
& \ \ +i\int d^4x F(x) t_L^{(0)}(x).\cr}\eqno(B-10)$$
The first term is identical to $I_{a2}$ and the second term exactly
cancells the contribution from diagram (7c).
There are still other contributions which are yet to be matched.
By inspection, one can easily show that the contributions from diagram (7b)
are precisely equal to those given by diagram (7d) with $t_R^{(1)}$ taken
to be $h{\phi^+b_R\over \phi^0}$. We hence complete the matching between
two different background field expansions on diagrams with one tree-level
top-quark propagator. It is straightforward to generalize above arguments
to cases with arbitrary numbers of top-quark propagators.

\def\psibL{\bar{\psi}_L}
\def\bR{b_R}
\def\phic{\phi^c}
\def\phbph{\bar{\phi} \phi}
\def\phib{\bar \phi}
\def\Dllm{\overleftarrow {D_{\mu}}^{\dagger}}

\def\Drlm{\overrightarrow {D_{\mu}}}
\def\Drum{\overrightarrow {D^{\mu}}}
\def\Dum{D^{\mu}}
\def\Dlm{D_{\mu}}
\def\dum{\partial^{\mu}}
\def\dlm{\partial_{\mu}}
\def\Dumph{(\Dum\phi)}
\def\Dlmph{(\Dlm\phi)}
\def\phida{\phi^{\dagger}}
\def\Dumphda{(\Dum\phi)^{\dagger}}
\def\Dlmphda{(\Dlm\phi)^{\dagger}}
\def\sigmn{\sigma^{\mu\nu}}
\def\Dln{D_{\nu}}
\def\Dlnph{(\Dln\phi)}
\def\Dlnphda{(\Dln\phi)^{\dagger}}
\def\lnphph{{\rm ln}({\phbph\over\kappa^2})}
\def\osp2{{1\over 16 \pi^2}}

\def\lnH2{{\rm ln}(H^2)}
\noindent
{\bf APPENDIX C: VERTICES WITH TWO DERIVATIVES}
\smallskip
In this appendix we give our results for terms involving two
derivatives with either no or two $\gamma$-matrices between the fermion
operators (chirality changing terms).  Their coefficients are calculated via
diagrams as in  Fig.1,
where in addition one neutral boson is explicitly emitted from the
internal fermion line, or a pair of neutral or charged bosons from the
internal boson line. The subtraction applied is described in Section 3.
The terms have the following Lorentz structures:
$$\eqalignno{
  &d_1\>((\psibL \Dllm)\cdot\phic)\>(\Drum \bR)\>/\phbph\>,\cr
  &d_2\>((\psibL \Dllm)\cdot\phic)\>(\Drum
\bR)\>\lnphph\>/\phbph\>,\cr
  &d_3\>((\psibL \Dllm)\cdot(\Dum\phic))\>\bR\>/\phbph\>,\cr
  &d_4\>((\psibL \Dllm)\cdot(\Dum\phic))\>\bR\>\lnphph\>/\phbph\>,\cr
  &d_5\>(\psibL\cdot(\Dlm\phic))\>\>(\Drum\bR)\>/\phbph\>,\cr
  &d_6\>(\psibL\cdot(\Dlm\phic))\>\>(\Drum\bR)\>\lnphph\>/\phbph\>,\cr
  &d_7\>((\psibL \Dllm)\cdot\phic)\>\bR\>\>[\Dumphda\cdot\phi -
       \phida\cdot\Dumph]\>/(\phbph)^2\>,\cr
  &d_8\>((\psibL \Dllm)\cdot\phic)\>\bR\>\>[\Dumphda\cdot\phi -
       \phida\cdot\Dumph]\>\lnphph\>/(\phbph)^2\>,\cr
  &d_9\>(\psibL\cdot\phic)\>(\Drlm\bR)\>\>[\Dumphda\cdot\phi -
       \phida\cdot\Dumph]\>/(\phbph)^2\>,\cr
  &d_{10}\>(\psibL\cdot\phic)\>(\Drlm\bR)\>\>[\Dumphda\cdot\phi -
       \phida\cdot\Dumph]\>\lnphph\>/(\phbph)^2\>,&(C.1)\cr
  &d_{11}\>(\psibL\cdot(\Dlm\phic))\>\bR\>\>[\Dumphda\cdot\phi -
       \phida\cdot\Dumph]\>/(\phbph)^2\>,\cr
  &d_{12}\>(\psibL\cdot(\Dlm\phic))\>\bR\>\>[\Dumphda\cdot\phi -
       \phida\cdot\Dumph]\>\lnphph\>/(\phbph)^2\>,\cr
  &d_{13}\>(\psibL\cdot\phic)\>\bR\>\>(\Dlmphda\cdot\Dumph)\>/(\phbph)^2\>,\cr
  &d_{14}\>(\psibL\cdot\phic)\>\bR\>\>(\Dlmphda\cdot\Dumph)
         \>\lnphph\>/(\phbph)^2\>,\cr

&d_{15}\>(\psibL\cdot\phic)\>\bR\>\>(\Dlmphda\cdot\phi)\>(\Dumphda\cdot\phi)\>
    /(\phbph)^3\>,\cr
  &d_{16}\>(\psibL\cdot\phic)\>\bR\>\>(\Dlmphda\cdot\phi)\>
    \lnphph\>/(\phbph)^3\>,\cr
  &d_{17}\>(\psibL\cdot\phic)\>\bR\>\>(\phida\cdot\Dlmph)\>(\phida\cdot\Dumph)
    \>/(\phbph)^3\>,\cr
  &d_{18}\>(\psibL\cdot\phic)\>\bR\>\>(\phida\cdot\Dlmph)\>(\phida\cdot\Dumph)
    \>\lnphph\>/(\phbph)^3\>,\cr
  &d_{19}\>(\psibL\cdot\phic)\>\bR\>\>(\Dlmphda\cdot\phi)\>(\phida\cdot\Dumph)
    \>/(\phbph)^3\>,\cr
  &d_{20}\>(\psibL\cdot\phic)\>\bR\>\>(\Dlmphda\cdot\phi)\>(\phida\cdot\Dumph)
    \>\lnphph\>/(\phbph)^3\> + H.c.\cr}$$

$$\eqalign{
  &e_1\>(\psibL\cdot i \sigmn (\Dlm\phic))\>(\Dln\bR)\>/\phbph\>,\cr
  &e_2\>(\psibL\cdot i \sigmn
(\Dlm\phic))\>(\Dln\bR)\>\lnphph\>/\phbph\>,\cr
  &e_3\>(\psibL\cdot i \sigmn \phic)\>(\Dlm\bR)\>\>
       [\Dlnphda\cdot\phi - \phida\cdot\Dlnph]\>/(\phbph)^2\>,\cr
  &e_4\>(\psibL\cdot i \sigmn \phic)\>(\Dlm\bR)\>\>
       [\Dlnphda\cdot\phi -
\phida\cdot\Dlnph]\>\lnphph\>/(\phbph)^2\>,\cr
  &e_5\>(\psibL\cdot i \sigmn (\Dlm\phic))\>\bR\>\>
       [\Dlnphda\cdot\phi - \phida\cdot\Dlnph]\>/(\phbph)^2\>,\cr
  &e_6\>(\psibL\cdot i \sigmn (\Dlm\phic))\>\bR\>\>
       [\Dlnphda\cdot\phi -
\phida\cdot\Dlnph]\>\lnphph\>/(\phbph)^2\>,\cr
  &e_7\>(\psibL\cdot i \sigmn \phic)\>\bR\>\>
       (\Dlmphda\cdot\Dlnph)\>/(\phbph)^2\>,\cr
  &e_8\>(\psibL\cdot i \sigmn \phic)\>\bR\>\>
       (\Dlmphda\cdot\Dlnph)\>\lnphph\>/(\phbph)^2\>,\cr
  &e_9\>(\psibL\cdot i \sigmn \phic)\>\bR\>\>
        (\Dlmphda\cdot\phi)\>(\phida\cdot\Dlnph)\>/(\phbph)^3\>,\cr
  &e_{10}\>(\psibL\cdot i \sigmn \phic)\>\bR\>\>
        (\Dlmphda\cdot\phi)\>(\phida\cdot\Dlnph)\>\lnphph\>/(\phbph)^ 3\>
        + H.c.\cr
}\eqno(C.2)
$$
Our aim is to determine the d's and the e's.  A few remarks are in order about
our choice of the operator basis.
Because of the special properties of the charge conjugated doublet $\phic$
in SU(2), and because of  the
completeness relation for the Pauli matrices, the number of
independent structures has been vastly reduced. Making use of the constraint
$\psibL \cdot  \phi = 0$
leads to our classification of having only products $\psibL \cdot
\phic$
and derivatives thereof. In the case of structures with $\sigma_{\mu  \nu}$,
the antisymmetry with respect to the Lorentz indices allows even  further
reduction to terms without derivatives on $\psibL$.  Note that we have not
included structures containing explicitly the field strengths $F_{\mu \nu}$
and $G_{\mu \nu}$.  They will lead to terms with three derivatives (if we
use them to replace $\sigma_{\mu \nu}$) or to terms without chirality change.
Thus, they can be clearly distinguished from the parts given in this
appendix.  The (anti-)hermitean
form of the single derivative acting on the scalar doublets
$[\Dlnphda\cdot\phi - \phida\cdot\Dlnph]$ is obtained by using
relations like
$$\eqalign{
(\dlm\psibL\cdot\phic)\>\bR\>(\dum\phib\cdot\phi) =
  &(\dlm\psibL\cdot\dum\phic)\>\bR\>(\phib\cdot\phi)
+ (\psibL\cdot\dlm\phic)\>\bR\>(\phib\cdot\dum\phi)\cr
&- (\psibL\cdot\phic)\>\bR\>(\dum\phib\cdot\dlm\phi)\>,
}\eqno(C.3)
$$
which are again consequences of the constraint. Similar
considerations
lead to the expectation that the coefficients $e_3$ and $e_4$ both
should vanish, which is verified below.

\noindent
The final results for the coefficients $d_i$ and $e_i$ are:
$$\eqalign{
   &d_1 = \osp2 (\>{1\over 2}\>h\>)\>, \ \
   d_2 = \osp2 (\>0\>)\>,\ \
   d_3 = \osp2 (\>h\>)\>,\ \
   d_4 = \osp2 (\>0\>)\>,\cr
   &d_5 = \osp2 ({1\over 12}h)\>,\ \
   d_6 = \osp2 (\>0\>)\>,\ \
   d_7 = \osp2 (\>{1\over 3}\>h\>)\>,\ \
   d_8 = \osp2 (\>0\>)\>,\cr
   &d_9 = \osp2 (\>{1\over 3}\>h\>)\>,\ \
   d_{10} = \osp2 (\>0\>)\>,\ \
   d_{11} = \osp2 (\>{1\over 3}\>h\>)\>,\ \
   d_{12} = \osp2 (\>0\>)\>,\cr
   &d_{13} = \osp2 (h({1\over 2 \epsilon'}-{1\over 2}\lnH2+{1\over
12}))\>,\ \
   d_{14} = \osp2 (-{1\over 2} h)\>,\ \
   d_{15} = \osp2 ({1\over 4} h)\>,\cr
   &d_{16} = \osp2 (\>0\>)\>,\ \
   d_{17} = \osp2 ({1\over 4} h)\>,\ \
   d_{18} = \osp2 (\>0\>)\>,\cr
   &d_{19} = \osp2 (h(-{1\over 2 \epsilon'}+{1\over 2}\lnH2+{1\over
6}))\>,\ \
   d_{20} = \osp2 ({1\over 2} h)\>,\cr
}\eqno(C.4)$$

$$\eqalign{
   &e_1 = \osp2 ({1\over 2} h)\>,\ \
   e_2 = \osp2 (\>0\>)\>,\ \
   e_3 = \osp2 (\>0\>)\>,\ \
   e_4 = \osp2 (\>0\>)\>,\cr
   &e_5 = \osp2 (-{1\over 4} h)\>,\ \
   e_6 = \osp2 (\>0\>)\>,\ \
   e_7 = \osp2 (h({1\over 2 \epsilon'}-{1\over 2}\lnH2+{5\over
6}))\>,\cr
   &e_8 = \osp2 (-{1\over 2} h)\>,\ \
   e_9 = \osp2 (-h({1\over 2 \epsilon'}-{1\over 2}\lnH2+{13\over
6}))\>,\ \
   e_{10} = \osp2 ({1\over 2} h)\>,\cr
}\eqno(C.5)
$$
\endpage

\centerline{\bf REFERENCES}
\medskip

\item {1.}
 See, for example, Particle Data Group, Phys. Rev. D45, S1 (1992).
\item {2.}
H. Steger, E. Flores and Y.-P. Yao, Phys. Rev. Lett. 59, 385 (1987);
G.-L. Lin, H. Steger and Y.-P. Yao, Phys. Rev. D 44, 2139, (1991).
\item {3.}
T. Appelquist and C. Bernard, Phys.
Rev. D 23, 425 (1981); A. C. Longhitano, Nucl. Phys. B188, 118 (1981); R.
Akhoury and Y.-P. Yao, Phys. Rev. D 25, 1605 (1982).
\item {4.}
 C. P. Burgess and D. London, McGill-92-05.

\item {5.}
 S. Coleman and E. Weinberg, Phys. Rev. D 7, 1888 (1973);
A. D. Linde, Sov. Phys. JETP Lett. 23, 64 (1976); S. Weinberg, Phys.
Rev. Lett. 36, 294 (1976).

\item {6.}
 M.
Veltman, Nucl. Phys. B123, 89 (1977); M. S. Chanowitz, M. A. Furman,
and I. Hinchliffe, Phys. Lett. 78B, 285 (1978); M. B. Einhorn, D. R.
T. Jones and M. Veltman, Nucl. Phys. B191, 146 (1981).

\item {7.}
 As it is very likely that $m_H>m_t$, we shall
restrict ourselves to the case where the Higgs boson $\sigma $ appears
only in the intermediate state.
\item {8.}
G.
Eilam, Phys. Rev. Lett. 49, 1478
(1982); W.-S. Hou, A. Soni and H. Steger, Phys. Rev. Lett. 59,
1521 (1987); W.-S. Hou, Nucl. Phys. B308, 561 (1988); R. Grigianis, P.
J. O'Donnell, M. Sutherland and H. Navelet, Phys. Lett. 224B, 209 (1989).

\item {9.}
B. Grinstein, R.
Springer and M. B. Wise, Phys. Lett. 202B, 138 (1988); W. S. Hou, A.
Soni and H. Steger, Phys. Lett. 192B, 441 (1987); J. L. Hewett, Phys.
Lett. 193B, 327 (1987); S. Bertolini, F. Borzumati and A. Masiero,
Phys. Lett. 192B, 437 (1987)
\splitout
N. G. Deshpande, G. Eilam, A. Soni and G. L. Kane, Phys. Rev. Lett. 57,
1106 (1986);
W.-S. Hou, R. S. Willey and A. Soni, Phys. Rev. Lett. 58, 1608 (1987);
B. Grinstein and M. B. Wise, Phys. Lett. 201B, 274 (1988);
W.-S. Hou and R. S. Willey, Phys. Lett. 202B, 591 (1988).
\item {10.}
T. Inami and C.S. Lim, Prog. Theor.
Phys. 65, 297 (1981).

\item {11.}
A. A. Akhundov, D. Yu.
Bardin and T. Riemann, Nucl. Phys. B276, 1 (1986); W. Beenakker and W.
Hollik, Z. Phys. C40, 141 (1988); J. Bernabeu, A. Pich and A.
Santamaria, Phys. Lett. B200, 569 (1988).

\item {12.}
W.A. Kaufman, H. Steger and Y.-P. Yao, Mod. Phys. Lett. A3,
1479 (1988).

\item {13.}
Low energy effects of heavy
quarks have also been condidered by F. Feruglio, A. Masiero and
L. Maiani, Nucl. Phys. B387, 523 (1992). We do not agree with the results
obtained in this article. For example, we have not been able to
use their tree Lagrangian to derive the corresponding result for $\phi^+
b_L$-scattering, as given by the full theory.

\item {14.}
W. Zimmermann,
Lectures on Elementary Particles and Quantum Field Theory, edited by
S. Deser, M. Grisaru and H. Pendleton (MIT Press, Cambridge, MA,
1970); Y. Kazama and Y.-P. Yao, Phys. Rev. D 25. 1605 (1982).

\item {15.}
{}From now on, any equation in Ref. [2] will be referred as Eq. (I-a.b) where
``a.b"  is the usual number denoting the equation.

\item {16.}
D. Yu. Bardin, B. M. Vilensky, P. Ch. Christova, Sov. J. Nucl. Phys. 53 (1991)
152; B. A. Kniehl, Nucl. Phys. B376 (1992) 3; A. Dabelstein and W. Hollik, Z.
Phys. C53 (1992) 507.

\endpage

\hskip -1cm\centerline{\bf FIGURE CAPTIONS}
\medskip

\item {Fig.1:}
One-loop diagrams contributing to $\bar b_L b_R$ self energy.
The blackened squares are subtractions of the appropriate tree
heavy vertices.
\bigskip
\item {Fig. 2:}
Tree diagrams contributing to $b-\phi ^+$ scattering.
\bigskip
\item {Fig. 3:}
One-loop diagrams contributing to $\bar b_Lb_L$ self energy.
\bigskip
\item {Fig. 4:}
One-loop diagram contributing to $\bar b_Rb_R$ self energy.
\bigskip
\item {Fig. 5:}
One loop diagrams contributing to $b_R-\phi ^+$ scattering.
\bigskip
\item {Fig. 6:}
Diagrams which illustrate the equivalence of using two different
background field solutions.

\end